\documentclass[]{raa}            

\usepackage{graphicx,times}             
\usepackage{natbib}
\usepackage{amssymb,amsmath}
\usepackage{pifont}
\usepackage{longtable}
\bibpunct{(}{)}{;}{a}{}{,}

\usepackage[pagebackref=true]{hyperref}

\begin{document}

  \title{Polycyclic Aromatic Hydrocarbons in the circumstellar medium of Herbig Ae/Be stars
}

   \volnopage{Vol.0 (20xx) No.0, 000--000}      
   \setcounter{page}{1}          

   \author{Roy Arun 
      \inst{1}
   \and Blesson Mathew
      \inst{2}
   \and Baskaran Shridharan
      \inst{3}
    \and Krishnan Ujjwal
      \inst{2}
    \and Krishna R Akhil
      \inst{2}
    \and Maheswar Gopinathan
      \inst{1}
    \and Sreeja S. Kartha
    \inst{2}
   }

   \institute{Indian Institute of Astrophysics, Sarjapur Road, Koramangala, Bangalore 560034, India {\it arunroyon@gmail.com}\\
        \and
             Center of Excellence in Astronomy and Astrophysics, Department of Physics and Electronics, CHRIST (Deemed to be University), Bangalore 560029, India\\
        \and
             Tata Institute of Fundamental Research, Homi Bhabha Road, Mumbai 400005, India\\
\vs\no
   {\small Received 20xx month day; accepted 20xx month day}}

\abstract{We present a comprehensive mid-infrared spectroscopic survey of 124 Herbig Ae/Be stars using newly processed \textit{Spitzer}/IRS spectra from the newly released CASSISjuice database. Based on prominent dust and molecular signatures (polycyclic aromatic hydrocarbons, silicates, and hydrogenated amorphous carbons), we classify the stars into five groups. Our analysis reveals that 64\% of the spectra show PAH emission, with detections peaking in the stellar effective temperature range 7000--11000\,K (B9--A5). Silicate features appear in 50\% of the sample and likewise diminish at higher temperatures. Additionally, we find that future PAH studies can focus on Herbig Ae/Be stars with a spectral index \(n_{2-24} > -1\) and flared morphologies to maximize PAH detections. The 6.2\,\(\mu\)m PAH band is the most frequently observed in our sample, shifting blueward with increasing stellar temperature, and this is the largest sample yet used to test that peak shift. The weaker 6.0\,\(\mu\)m feature does not shift with 6.2\,\(\mu\)m, implying a distinct origin of C=O (carbonyl) or olefinic C=C stretching relative to C--C vibrations. We examined the 11.0/11.2\,\(\mu\)m PAH ratio using high-resolution \textit{Spitzer} spectra for the first time in a sample of Herbig Ae/Be stars, finding a range of ionization conditions. This study provides a strong foundation for future \textit{JWST} observations of intermediate-mass pre-main-sequence stars.
\\$\cdots\cdots$
\keywords{techniques: spectroscopic --- stars: variables: T Tauri, Herbig Ae/Be --- astrochemistry --- catalogues}
}

   \authorrunning{R. Arun et al. }            
   \titlerunning{Polycyclic Aromatic Hydrocarbons in the circumstellar medium of Herbig Ae/Be stars}  

   \maketitle

%
%
\section{Introduction}           
\label{sect:intro}

Herbig Ae/Be (HAeBe) stars constitute an important population of intermediate-mass pre-main sequence (PMS) stars, bridging the gap between the well-studied low-mass T Tauri stars and the more massive, often heavily embedded young stellar objects. Their relatively short pre-main sequence lifetimes and intense UV radiation fields make them ideal laboratories for investigating the accretion processes, disk and chemical evolution \citep[e.g.,][]{Meeus2001A&A...365..476M, Fairlamb2017MNRAS.464.4721F, Vioque2018A&A...620A.128V}. A notable characteristic of many HAeBe stars is their mid-infrared (MIR) spectra, which, while not as molecule-rich as those seen in T Tauri disks (as revealed by recent JWST observations : \citealp{Grant2023ApJ...947L...6G,Temmink2024A&A...686A.117T}), prominently display emissions from dust grains (e.g., silicates) and from polycyclic aromatic hydrocarbons (PAHs). The relatively strong UV radiation from HAeBe stars is sufficient to excite PAH emission, providing a valuable opportunity to study the evolution of PAHs into other complex organic molecules during critical phases of planet formation.

Previous ground- and space-based studies have highlighted the diversity of molecular features found in HAeBe stars. Low-mass T Tauri stars, for instance, display significant silicate emission, whereas HAeBe stars often exhibit prominent PAH signatures, reflecting their higher UV flux and, in some cases, more flared disk structures \citep{Acke2010, Juhasz2010ApJ...721..431J, Seok2017ApJ...835..291S}. In the interstellar medium (ISM), PAHs are widespread tracers of photodissociation regions (PDRs) and can exist in various states of ionization, morphology, and size \citep{Tielens2008ARA&A..46..289T}. However, their resilience and evolution in the harsher environments surrounding earlier-type HAeBe stars remain open questions, with evidence suggesting that extreme UV radiation may destroy or heavily process these molecules \citep[e.g.,][]{Habart2004A&A...427..179H, Maaskant2013A&A...555A..64M}. Equally intriguing are silicate grains, which are abundant in protoplanetary disks yet may be transformed—or even depleted—in the presence of strong UV fields \citep{Meeus2001A&A...365..476M, Bouwman2001A&A...375..950B}.

The unprecedented volume of mid-infrared spectra from the \textit{Spitzer} Infrared Spectrograph (IRS) archive provides a powerful resource for revisiting and expanding our knowledge of HAeBe stars. Recent developments in data reduction pipelines—specifically, the publicly released CASSISjuice package \citep{CASSIS2011ApJS..196....8L,Lebouteiller2015ApJS..218...21L,CASSISJUICE2023arXiv230906876L}—enable a more refined and comprehensive assessment of these targets. By systematizing and uniformly processing both low- and high-resolution \textit{Spitzer}/IRS staring-mode spectra, the CASSISjuice dataset opens new avenues for large-sample statistical analyses. For instance, the \textit{Spitzer} Spectral HAeBe Catalog (SSHC) \citep{Arun2023MNRAS.523.1601A} forms the basis of this work, hosting hundreds of spectra spanning a wide parameter space in stellar effective temperature, mass, and disk geometry.

PAH and silicate emissions serve as critical diagnostics of disk chemistry and dust processing, offering insights into the mechanisms governing planet formation timelines. In contrast to earlier studies that primarily focused on individual HAeBe stars or limited samples, our work leverages a uniformly processed dataset to deliver a more comprehensive statistical analysis. Specifically, we aim to classify the molecular features in HAeBe stars and examine their relationship with stellar properties, thereby enhancing our understanding of disk evolution processes. The sample data are described in Section 2. We present our analysis and results in Section 3, including a classification of the stars based on the presence or absence of key features such as PAHs and silicates. We then investigate how the detection frequencies of these features correlate with stellar and disk parameters—most notably the effective temperature and the continuum spectral index. Finally, in Section 5, we summarize our key findings and discuss their implications for future studies.
\section{Data}

\subsection{\textit{Spitzer} IRS Catalog of HAeBe Stars}

The data utilized in this study originates from the SSHC detailed in \citet{Arun2023MNRAS.523.1601A}. This catalog plays a pivotal role in our exploration of molecular and dust features around HAeBe stars. Notably, the SSHC led to the identification of C\textsubscript{60} vibrational modes at 17.4 and 18.9 $\mu$m in nine HAeBe stars, establishing its utility for broader scientific applications.

The catalog comprises low- and high-resolution spectra of 124 HAeBe stars, sourced primarily from \citet{Vioque2018A&A...620A.128V}. The high-resolution spectra reveal various metallic and gas lines, including both allowed and forbidden transitions, and can be compared with the spectral features of low-mass T-Tauri stars \citep{Akhila2025MNRAS.tmp..903A}. This paper aims to classify stars based on the presence of molecular features such as PAHs, silicates, and fullerenes, positioning this work as a foundation for future JWST observations of HAeBe stars.

Spectra were retrieved from the Combined Atlas of Sources with \textit{Spitzer} IRS Spectra (CASSIS\footnote{CASSIS is a product of the IRS instrument team, supported by NASA and JPL.}) archive \citep{CASSIS2011ApJS..196....8L,Lebouteiller2015ApJS..218...21L}. Low-resolution IRS (SL/LL) spectra cover a wavelength range of 5--38 $\mu$m for 55 stars, while for 39 stars, the coverage is 5--15 $\mu$m. High-resolution IRS (SH/LH) spectra, ranging from 10--20 $\mu$m, are available for 79 stars. For 47 stars, both low- and high-resolution spectra are available, providing a comprehensive data set for molecular and dust feature analysis. More details on the catalog are available in \citet{Arun2023MNRAS.523.1601A}.

\subsection{CASSISJuice}

CASSISjuice is an enhanced and offline version of the CASSIS pipeline and atlas, designed to handle reduced spectra from the \textit{Spitzer} Infrared Spectrograph (IRS), which is released in September 2023 \citep{CASSIS2011ApJS..196....8L,Lebouteiller2015ApJS..218...21L,CASSISJUICE2023arXiv230906876L}. It was created with a primary objective to facilitate and optimize the utilization of archival \textit{Spitzer}/IRS observations and complement forthcoming data from the JWST and other infrared facilities. Unlike CASSIS, CASSISjuice incorporates an extended set of previously unreleased spectra, making it a more comprehensive resource for the community. It offers two crucial components: an upgraded low- and high-resolution pipeline that processes all IRS staring-mode observations, and a complete atlas with an enhanced organization of pointings, resulting in a single CASSISjuice ID for each targeted position. This streamlined dataset can be accessed and explored through public repositories, which are hosted on Zenodo and GitLab\footnote{https://gitlab.com/cassisjuice} currently. We have  made use of the CASSISJuice to organize and analyse the SSHC in this work. 

The best extraction spectra is taken for low resolution and full aperture extraction is taken for high resolution spectra from CASSISjuice. Full aperture extraction is chosen for high resolution spectra as we are studying the molecular features present in the spectra. In the case of high-resolution products, CASSISjuice has the capability to disentangle the point-source emission from the extended physical background in the observation\footnote{https://irs.sirtf.com/Smart/CassisHRPipeline}. The full aperture extraction method co-adds the pixels in the detector corresponding to the particular wavelength value to compute the flux while bad pixels are identified and replaced.

\section{Analysis and Results}

We analyzed \textit{Spitzer} IRS spectra of 124 HAeBe stars from the SSHC, which has not previously been studied as a statistical sample. Our analysis involved classifying the spectra based on the molecular species present, following a systematic approach outlined below.

\subsection{Classification of \textit{Spitzer} IRS Spectra of HAeBe Stars}

We identified two dominant molecular species in the IRS spectra---PAHs and silicates---consistent with earlier studies \citep{Acke2010, Juhasz2010ApJ...721..431J}. While C\textsubscript{60} features at 17.4 and 18.9\,$\mu$m were previously detected in nine stars from this sample \citep{Arun2023MNRAS.523.1601A}, they are not included as a classification category in this study. In contrast, we tagged the two stars found with Hydrogenated Amorphous Carbons (HACs) by \cite{Arun2023MNRAS.523.1601A} in the 5--10\,$\mu$m region, HD~319896 and SAO~220669, as a separate category because they do not show PAH or silicate signatures in their spectra.

The most prominent molecular features are concentrated in the $5$--$15~\mu$m wavelength region, which includes the key PAH and silicate bands. Therefore, we based our classification on the spectral characteristics within this range, focusing particularly on the $6.2$, $7.7$, $8.6$, $11.2$, and $12.7~\mu$m PAH features and the $9.7~\mu$m silicate feature.

We visually classified the stars into five distinct categories: PAH only (P), silicate only (S), PAH \& Silicates (PS), No feature (NF), and HACs. This scheme builds on the classification system proposed by \cite{Boersma2008A&A...484..241B}, with the addition of the NF class to account for stars lacking PAH or silicate features. Importantly, NF does not imply an absence of spectral lines — HD 76534, for example, shows strong gas emission lines and forbidden transitions despite being classified as NF. Another similar classification scheme was adopted by \cite{Cerrigone2009ApJ...703..585C} for young planetary nebulae and post-AGB stars. They also did not have classification for sources without silicates and PAHs.

For the 31 stars with only high-resolution spectra (SH/LH; starting at 10\,$\mu$m), we employed a different strategy. Since the 5–10\,$\mu$m region is absent in these spectra, we relied on prominent mid-IR features such as the 11.2 and 16.4\,$\mu$m PAH bands, the red wing of the 9.7\,$\mu$m silicate feature, and the broader 20\,$\mu$m silicate bump.

The classification scheme for the SSHC is summarized as follows:\\

\noindent For SL/LL spectra 

\begin{itemize} 
\item PAH (P): Stars with one or more PAH emission features but no evidence of the $9.7~\mu \mathrm{m}$ silicate feature. 
\item Silicate (S): Stars exhibiting a prominent, broad $9~\mu \mathrm{m}$ silicate emission feature without accompanying PAH features. 
\item PAH \& Silicates (PS): Stars displaying both PAH and silicate features, an extension of the scheme used in previous literature \citep{Boersma2008A&A...484..241B}. 
\item No feature (NF): Stars lacking both PAH features and the $9.7~\mu \mathrm{m}$ silicate feature. Notably, some of these spectra contain strong gas emission lines. 
\item HACs: Two stars showing distinct spectral characteristics, including signatures of HACs in the $5-10~\mu \mathrm{m}$ region. 
\end{itemize}

\noindent For SH/LH spectra

\begin{itemize}
    \item PAH (P): Stars showing prominent PAH emission at 11.2\,$\mu$m and/or 16.4\,$\mu$m.
    \item Silicate (S):Stars with a declining continuum starting at 10\,$\mu$m and exhibiting the red wing of the broad 9.7\,$\mu$m silicate feature, along with a silicate bump peaking at 20\,$\mu$m.
    \item PAH  \& Silicate (PS): Stars displaying both PAH features (11.2 and/or 16.4\,$\mu$m) superimposed on the silicate continuum described above.
    \item No Features (NF): Spectra lacking identifiable PAH or silicate features, showing only atomic lines.
\end{itemize}

 We classified 46 stars as ``PS," 33 as ``P," 27 as ``NF," and 14 as ``S". Two stars were categorized as "HACs" due to unique spectral characteristics in the $5-10 \mu \mathrm{m}$ region. Additionally, we marked two spectra as bad (``B,") indicating data inconsistencies or artifacts. Interestingly, we do not find any sources with Silicates in absorption. No other sources are identified with HACs in their spectra other than two sources from \cite{Arun2023MNRAS.523.1601A}. Our sample comprises 79 spectra with PAH detections ("P" and "PS") and 43 spectra without PAH features, avoiding two ``B", resulting in a 64\% PAH detection frequency, consistent with previous findings for Herbig stars \citep{Seok2017ApJ...835..291S}. A closer examination of the ``P'' and ``PS'' sources reveals that the 6.2\,$\mu$m PAH feature is the most frequently observed in our sample. This high detection rate is largely influenced by observational biases. In particular, the 6.2\,$\mu$m feature is relatively isolated from strong dust continuum features, unlike the 7.7, 8.6, and 11.2\,$\mu$m bands which are frequently blended with, broad silicate emission in ``PS'' sources. While continuum subtraction using spline fitting is possible, it becomes increasingly unreliable in these blended regions. As a result, the 6.2\,$\mu$m band is more easily and confidently detected in both ``P'' and ``PS'' classes, making it the most frequently identified PAH feature in our sample.

\begin{figure}
    \centering
    \includegraphics[width=0.49\columnwidth]{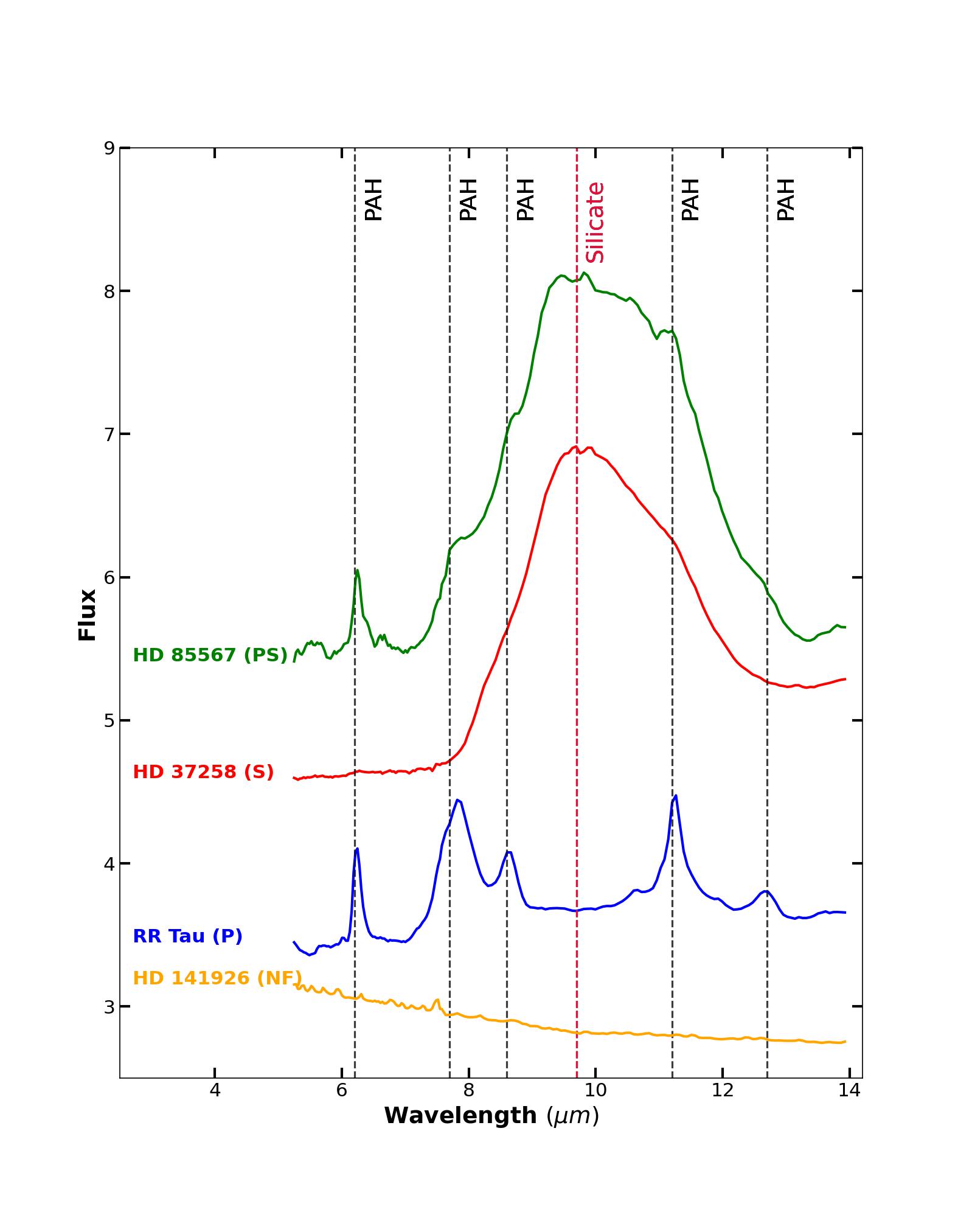}
    \includegraphics[width=0.49\columnwidth]{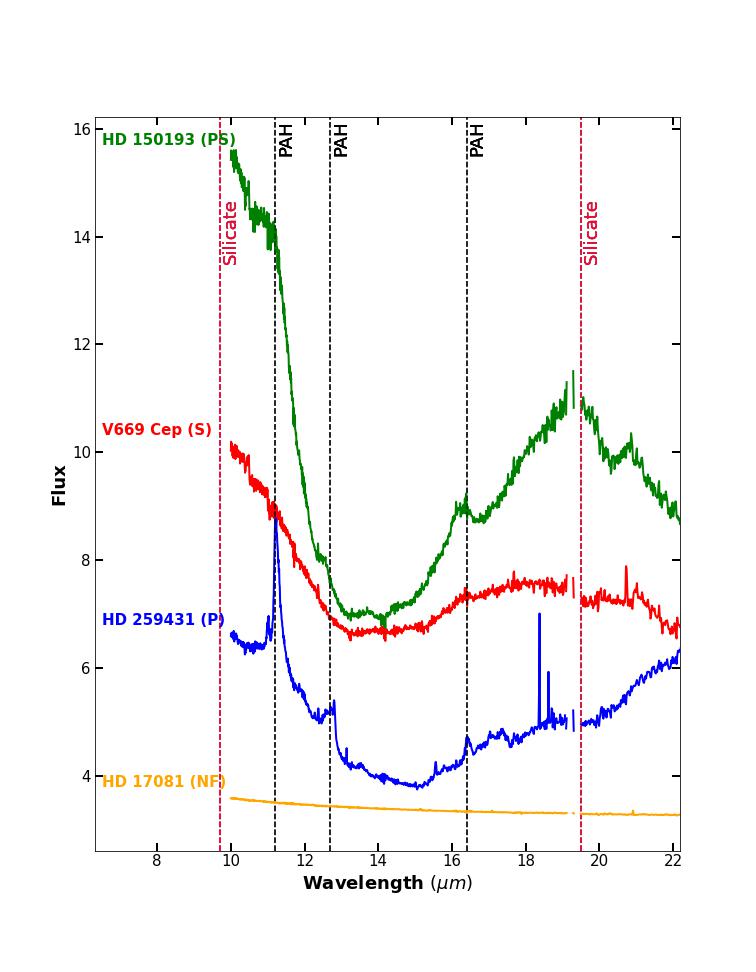}

    \caption{Representative examples of \textit{Spitzer} IRS spectra for HAeBe stars, categorized according to their molecular features. The left panel displays low-resolution spectra and the right panel shows high-resolution spectra, with fluxes plotted in arbitrary units for clarity. Color coding denotes the classification: blue for P, red for S, green for PS, and yellow for NF. Major PAH and silicate features are denoted with dashed lines. (Note: The two stars classified as HACs are not included here; see Figure 5 in \cite{Arun2023MNRAS.523.1601A} for details.)}
    \label{fig:class}
\end{figure}

\cite{Meeus2001A&A...365..476M} introduced a classification scheme for HAeBe stars based on the shape of their SEDs. In this scheme, Group I sources exhibit a rising mid- to far-infrared SED, indicative of flared disk geometries, whereas Group II sources show a more subdued SED, characteristic of self-shadowed disks. The classification of the IRS spectra were further examined in terms of their distribution across Meeus Groups I and II. Notably, we found that sources in Group I exhibit a higher frequency of PAH detections compared to Group II, consistent with the trend that flared disk geometries are more conducive to PAH excitation \citep{Meeus2001A&A...365..476M, Acke2004A&A...426..151A, Habart2004A&A...427..179H}. Specifically, 61\% of the "PS" and "P" sources belonged to Group I, which aligns with findings in the literature suggesting that PAHs are preferentially excited in flared disks due to increased UV exposure. The statistics of the classification and CASSIS Juice IDs are given in Table \ref{tab:table1}. The different classes of spectra identified in the study are shown in Figure. \ref{fig:class}.

\subsection{Spectral Index Distribution}

HAeBe stars are actively accreting matter from their gas- and dust-rich circumstellar disks \citep{VIEIRA2003,Vioque2018A&A...620A.128V}. Yet, within our sample, a subset of HAeBe stars shows a striking absence of both PAH and silicate emission features. For silicates, this absence has been linked to disk gaps, potentially indicating significant disk-clearing processes \citep{Maaskant2013A&A...555A..64M} or advanced grain growth. The lack of PAH features is particularly intriguing, given that PAHs are typically detected in UV-rich environments such as planetary nebulae, reflection nebulae, and photodissociation regions. We explore whether this could be due to UV photon-induced processing or dissociation during disk evolution, which may play a crucial role in the depletion of PAHs in these environments.

\begin{figure*}[]
    \centering
    \includegraphics[width=0.49\columnwidth]{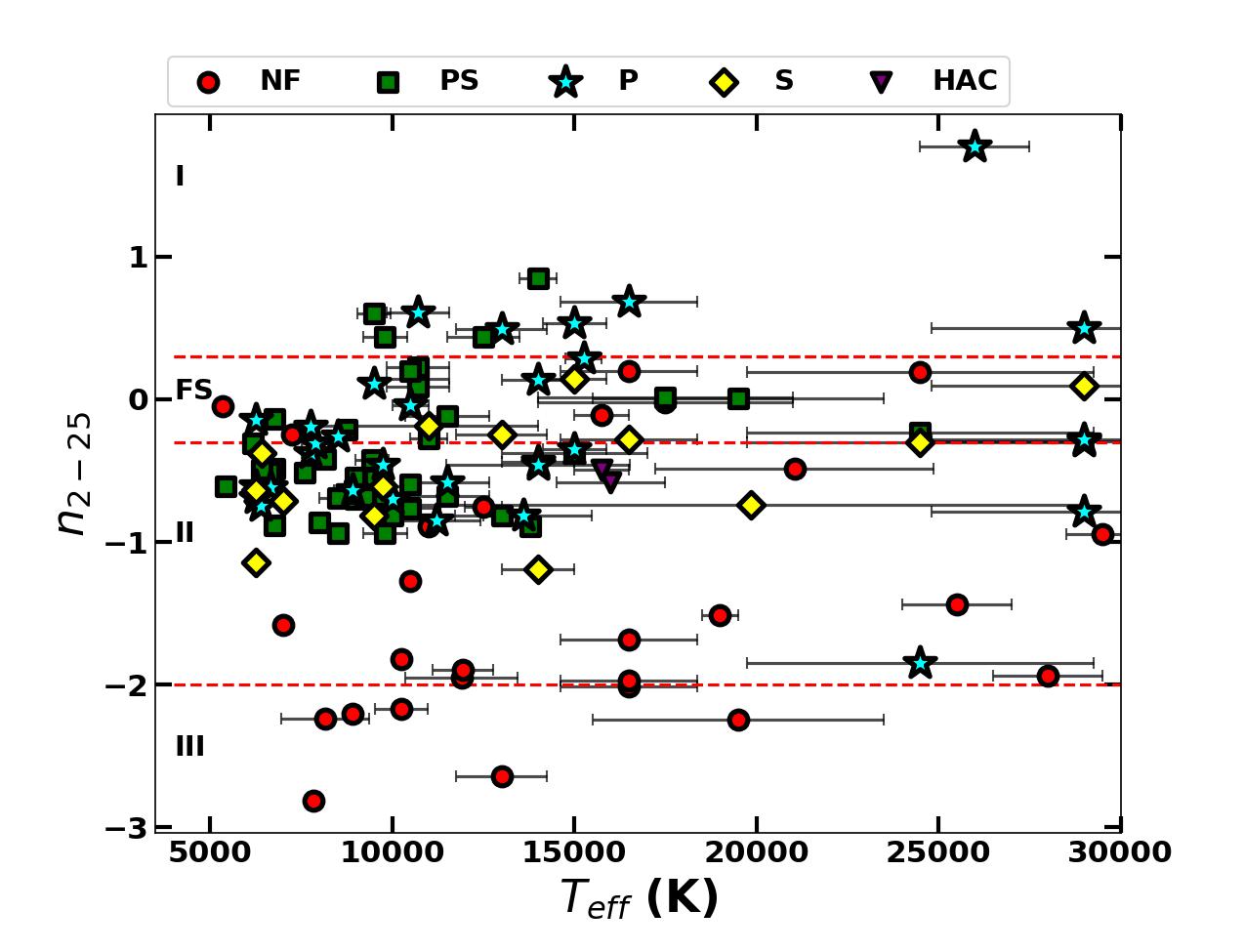}
    \includegraphics[width=0.49\columnwidth]{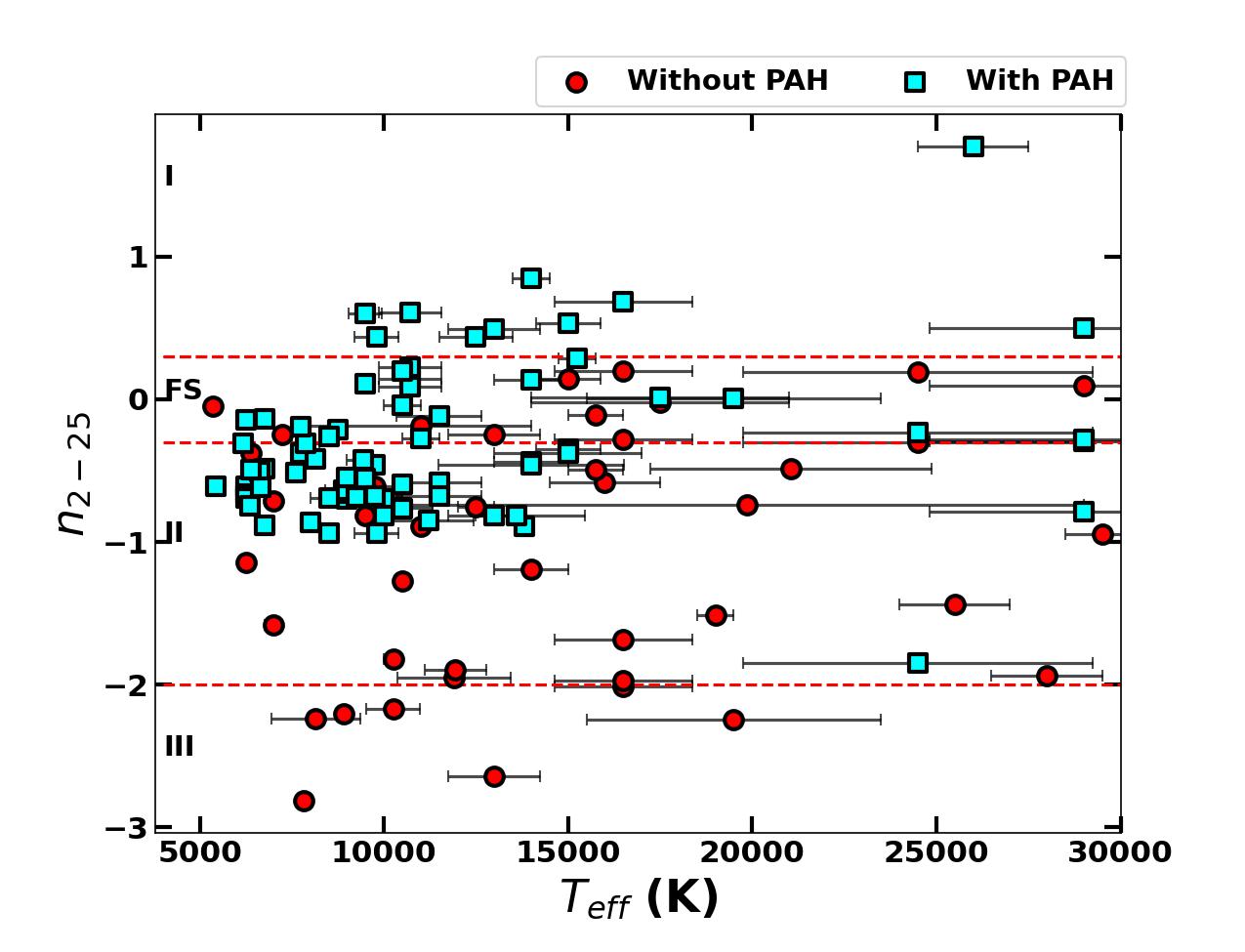}
    \caption{Distribution of the continuum spectral index ($n_{2-24}$) versus stellar effective temperature (T\textsubscript{eff}) for the sample of HAeBe stars analyzed in this study. The left panel is color-coded according to the spectral classification from the IRS spectra, highlighting the different evolutionary stages (Class I, Flat Spectrum, Class II, Class III). The right panel distinguishes stars with PAH emission (red) from those without PAH features (blue). Stars with $n_{2-24} > -1$ predominantly show PAH features, while those with $n_{2-24} < -1$ generally lack PAH detections.}
    \label{fig:ind_teff}
\end{figure*}

To address these questions, we conducted an analysis of various star and disk parameters. We calculated the continuum spectral index, denoted as $n_{2-24}$, for all the HAeBe stars in our sample. This index is widely used to classify young stellar objects (YSOs) into categories such as Class I, Flat Spectrum (FS), Class II, and Class III sources, providing a diagnostic of the disk evolution and circumstellar environment \citep{LAda1987, Wilking1989, Green1994, Manoj2011ApJS..193...11M}. The index is defined as the slope of the log-linear SED between two wavelengths:

\begin{equation}
n_{2-24} = \frac{\log\left( \lambda_{24} F_{\lambda_{24}} \right) - \log\left( \lambda_{2} F_{\lambda_{2}} \right)}{\log(\lambda_{24}) - \log(\lambda_{2})}
\end{equation}

where \( \lambda_2 = 2.2\,\mu\mathrm{m} \) (2MASS \( K_s \)-band) and \( \lambda_{24} = 22\,\mu\mathrm{m} \) (WISE W4 band). The fluxes were corrected for interstellar extinction using the visual extinction values from \citet{Vioque2018A&A...620A.128V} and assuming an average extinction law with \( R_V = 3.1 \). We converted magnitudes to flux densities using standard zero-point calibrations. Based on the value of the spectral index \( n_{2-24} \), young stellar objects (YSOs) can be classified into distinct evolutionary stages that reflect the amount and geometry of circumstellar material. Class~I objects (\( n_{2-24} > 0.3 \)) are deeply embedded protostars surrounded by infalling envelopes and thick disks, resulting in rising mid-IR SEDs. Flat Spectrum sources (\( 0.3 \geq n_{2-24} \geq -0.3 \)) are in transition between embedded and disk-dominated phases. Class~II sources (\( -0.3 > n_{2-24} > -2 \)) are characterized by optically thick, gas-rich disks without envelopes, and produce strong IR excess due to reprocessing of stellar radiation in the disk surface layers such as HAeBe stars and T Tauri stars. Class~III objects (\( n_{2-24} \leq -2 \)) have largely dissipated their disks, with SEDs dominated by the stellar photosphere and weak/absent IR excess \citep{LAda1987,Manoj2011ApJS..193...11M, Arun2023MNRAS.523.1601A}. This classifications allows us to relate spectral index values to disk evolution, which might affect the the presence and absence of PAH and silicate emission features.

The distribution of $n_{2-24}$ relative to the stellar effective temperature (T\textsubscript{eff}) is illustrated in Figure \ref{fig:ind_teff}. Both $n_{2-24}$ and $T_{\mathrm{eff}}$ are derived from broadband photometry (2MASS and WISE) and stellar parameters, rather than from the \textit{Spitzer}/IRS spectra themselves. This allows us to evaluate trends in disk structure and stellar temperature independent of mid-IR spectroscopic coverage, and to potentially apply these diagnostics to a broader sample of HAeBe stars. The left side of the figure is color-coded based on IRS spectral classification, while the right side distinguishes stars with and without PAH emission features. 

The majority of stars in our sample are classified as Class II and FS sources, confirming their pre-main sequence (PMS) status. Figure \ref{fig:ind_teff} shows that Class II stars with $n_{2-24}$ greater than -1 tend to exhibit PAH emission, suggesting that objects with PAH features are clustered in specific regions within the continuum spectral index distribution. This clustering provides a useful guide for identifying stars with a high probability of PAH detection in future observations. Interestingly, the one star showing PAH emission (HD 130437) below an index value of -1 only exhibits faint 11.2 $\mu$m features in its high-resolution spectra. The $K_s$ (2.2\,$\mu$m) and WISE W4 (22\,$\mu$m) bands used to compute the $n_{2-24}$ index lie outside the wavelengths of the most prominent PAH emission features—3.3\,$\mu$m and 11.2\,$\mu$m respectively. The W4 band spans a broader range (20–28\,$\mu$m), where PAH emission is weak and the mid-IR flux is dominated by thermal emission from the disk. As such, PAH contamination in the $n_{2-24}$ index is minimal. The observed clustering of PAH-rich sources at $n_{2-24} > -1$ is therefore attributed to the underlying disk structure, with flared disks producing stronger mid-IR excess and offering a more favorable geometry for PAH excitation.

Despite the limited clustering of PAH-detected sources in Figure \ref{fig:ind_teff}, stars without PAH features are more scattered across the index distribution. This implies that the absence of PAHs around certain HAeBe stars cannot be solely attributed to the spectral index of the star. However, the frequency of PAH detections is notably higher in stars with an index value greater than -1. Recent surveys and techniques continue to identify new HAeBe stars \citep{Vioque2020A&A...638A..21V,Nidhi2023MNRAS.524.5166N, Zhang2022ApJS..259...38Z}, and approximately 50\% of the HAeBe star sample from \cite{Vioque2018A&A...620A.128V} has not been observed by \textit{Spitzer}. Thus, future observation campaigns aiming to study PAH and silicate features in intermediate-mass young stars could prioritize stars with $n_{2-24}$ values greater than -1 for optimal PAH detection.

\subsection{A ``Sweet Spot" of PAH Detection}

\begin{figure}[]
    \centering
    \includegraphics[width=0.49\columnwidth]{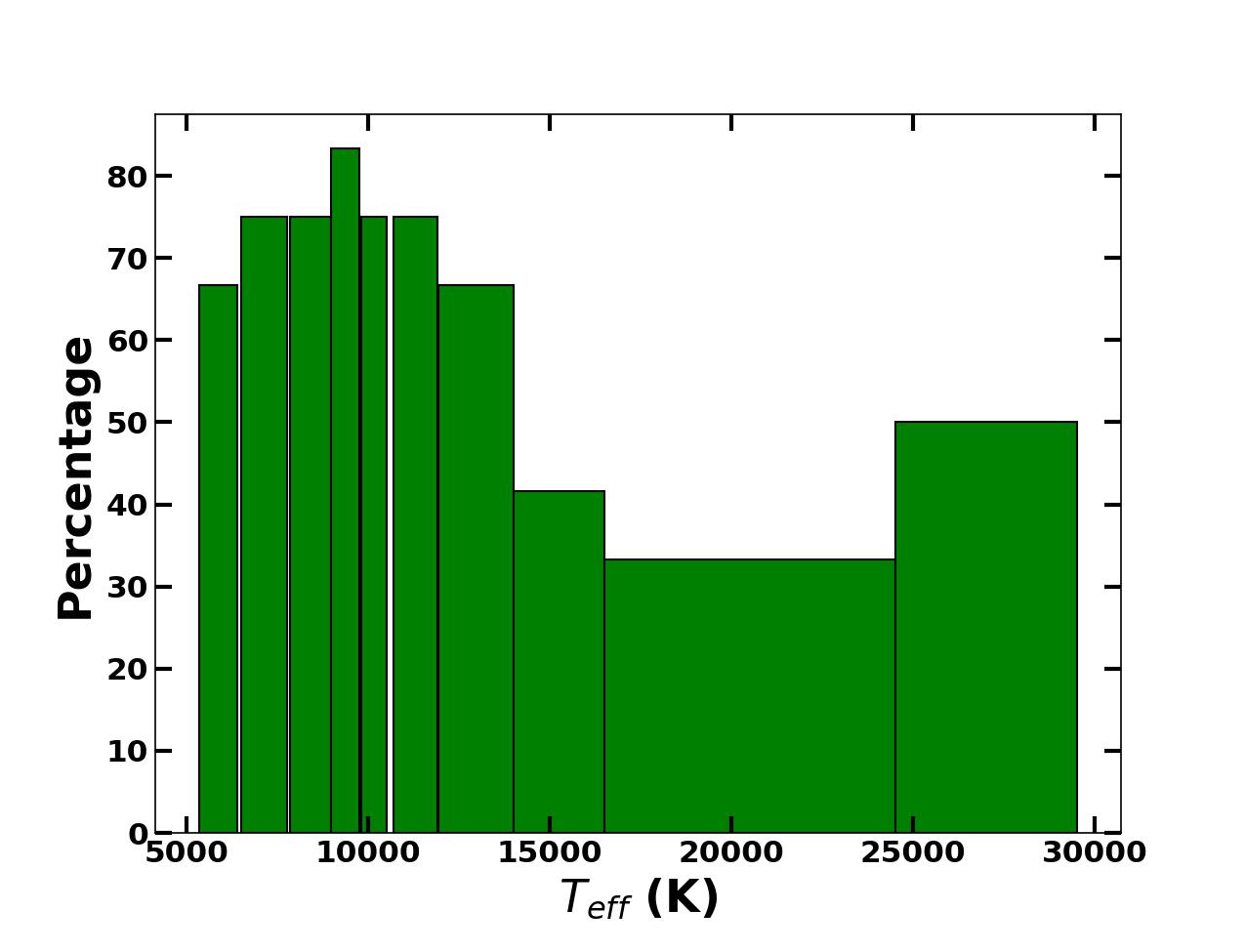}
    \includegraphics[width=0.49\columnwidth]{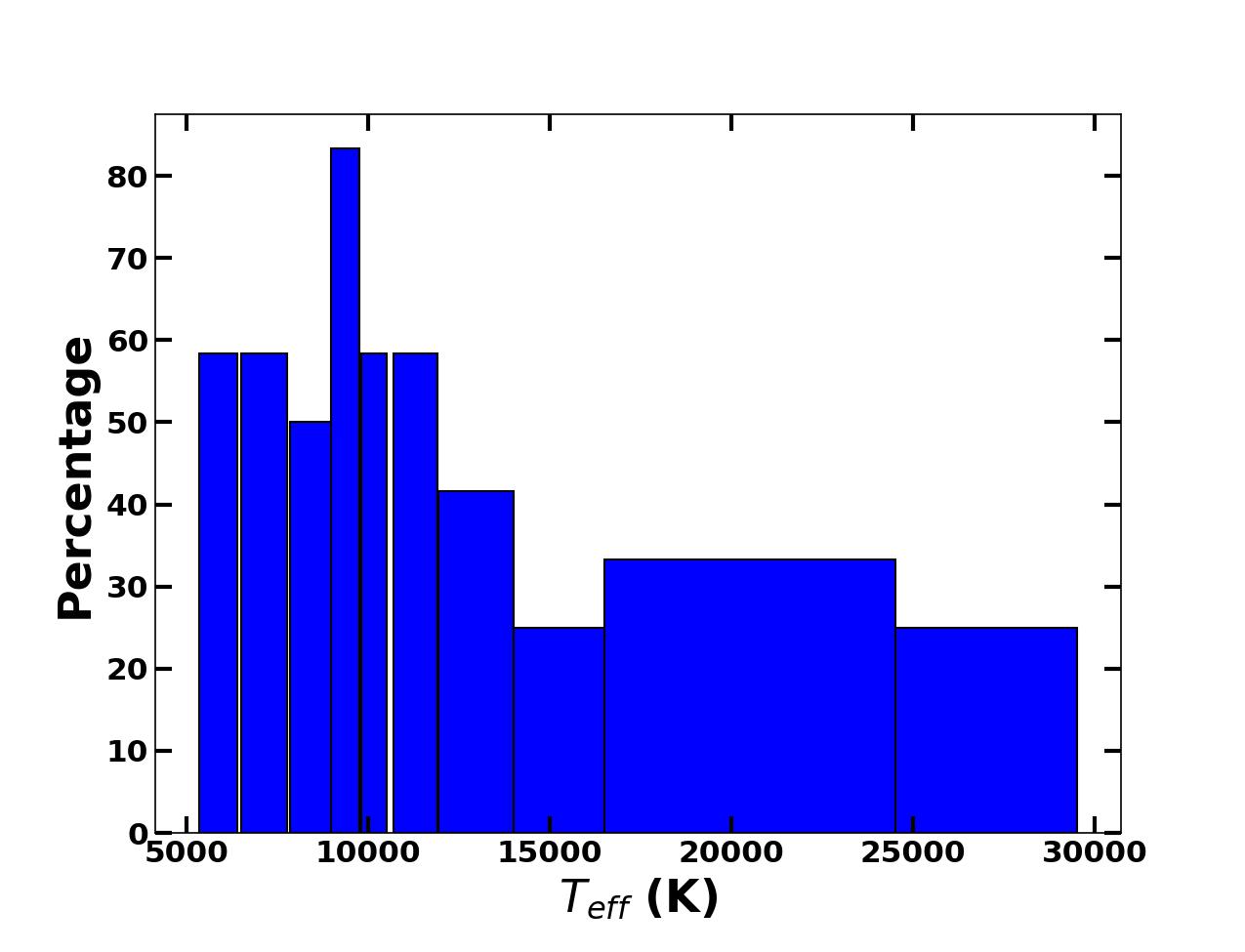}
    \caption{Detection frequency of PAH (left) and silicate (right) features as a function of effective temperature (T\textsubscript{eff}) for HAeBe stars. PAH detection shows a peak in cooler stars, declining sharply beyond T\textsubscript{eff} = 9360 K. Silicate features similarly peak around this temperature but maintain moderate detection rates ($\sim$55\%) up to T\textsubscript{eff} = 13,000 K, after which detection drops significantly, likely due to increased UV flux in hotter B-type stars.}
    \label{fig:detection}
\end{figure}

A clear detection frequency trend emerges when examining the T\textsubscript{eff} parameter in our sample with PAH detection. We computed the percentage of sources exhibiting PAHs across different T\textsubscript{eff} ranges. The sample was divided into 10 bins, each containing 12 stars. Figure \ref{fig:detection} presents a bar diagram of T\textsubscript{eff} versus the percentage of PAH detections.

The data show a gradual increase in the percentage of PAH detections from the bin centered around 5800 K up to 9360 K, followed by a steady decline toward 20,000 K. This indicates that PAH features are more prevalent at lower temperatures, while the detection frequency decreases significantly as T\textsubscript{eff} rises. In contrast, stars categorized as "No feature" exhibit an inverse relationship, with their numbers increasing with T\textsubscript{eff}. A slight rise in the final bin may be attributed to stars with associated nebulosity.

This trend suggests that the high energy output of early-type HBe stars, particularly their extreme UV photon flux, may play a role in the destruction or processing of PAHs in their circumstellar environments. When considering the detection frequency of PAHs in T Tauri and intermediate-mass T Tauri stars (approximately 10\%; \citealp{Acke2004A&A...426..151A, Geers2006A&A...459..545G}), it becomes clear that stars with T\textsubscript{eff} between 7000 K and 11,000 K, which corresponds to spectra range of B9 - F1, represent a favorable range for PAH detection. This temperature range seems to efficiently excite PAH vibrational modes, making it a ``sweet spot" for PAH detection in circumstellar disks.

Moreover, while PAHs are frequently destroyed or heavily processed around higher-temperature sources (e.g., B-type stars), fullerene features appear exclusively near such stars \citep{Arun2023MNRAS.523.1601A}, and are absent in A-type Herbig stars or later. This distinction supports the hypothesis of top-down fullerene formation, wherein PAHs are processed into fullerenes under harsher UV irradiation conditions \citep{Berne2015A&A...577A.133B}.

\subsection{Silicate Detection Frequency}

In a similar manner to the PAH detection analysis, we investigated the detection frequency of silicate features across different T\textsubscript{eff} ranges in our sample of HAeBe stars. The sample were divided into 10 bins, similar to previous section, to explore how silicate detection correlates with stellar effective temperature. The total detection frequency of silicates in our sample is 50\%, which is lower than PAHs (64\%). Also, we do not find any silicate absorption features in our sample. The bin diagram of silicate detection is shown in Figure \ref{fig:detection}.

The detection of silicates follows a distinct trend from PAHs. The highest detection rate occurs in the bin centered around 9360 K, where 83\% of the stars exhibit silicate features. But mostly the bins from 5875 to 13,000 K have similar detection rate of around 55\%, with a significant drop in detection frequency in the hottest stars. For instance, only 25\% of stars in the bin centered around 27000 K show detectable silicate features.
\begin{figure}[htbp]
    \centering
    \includegraphics[width=0.32\columnwidth]{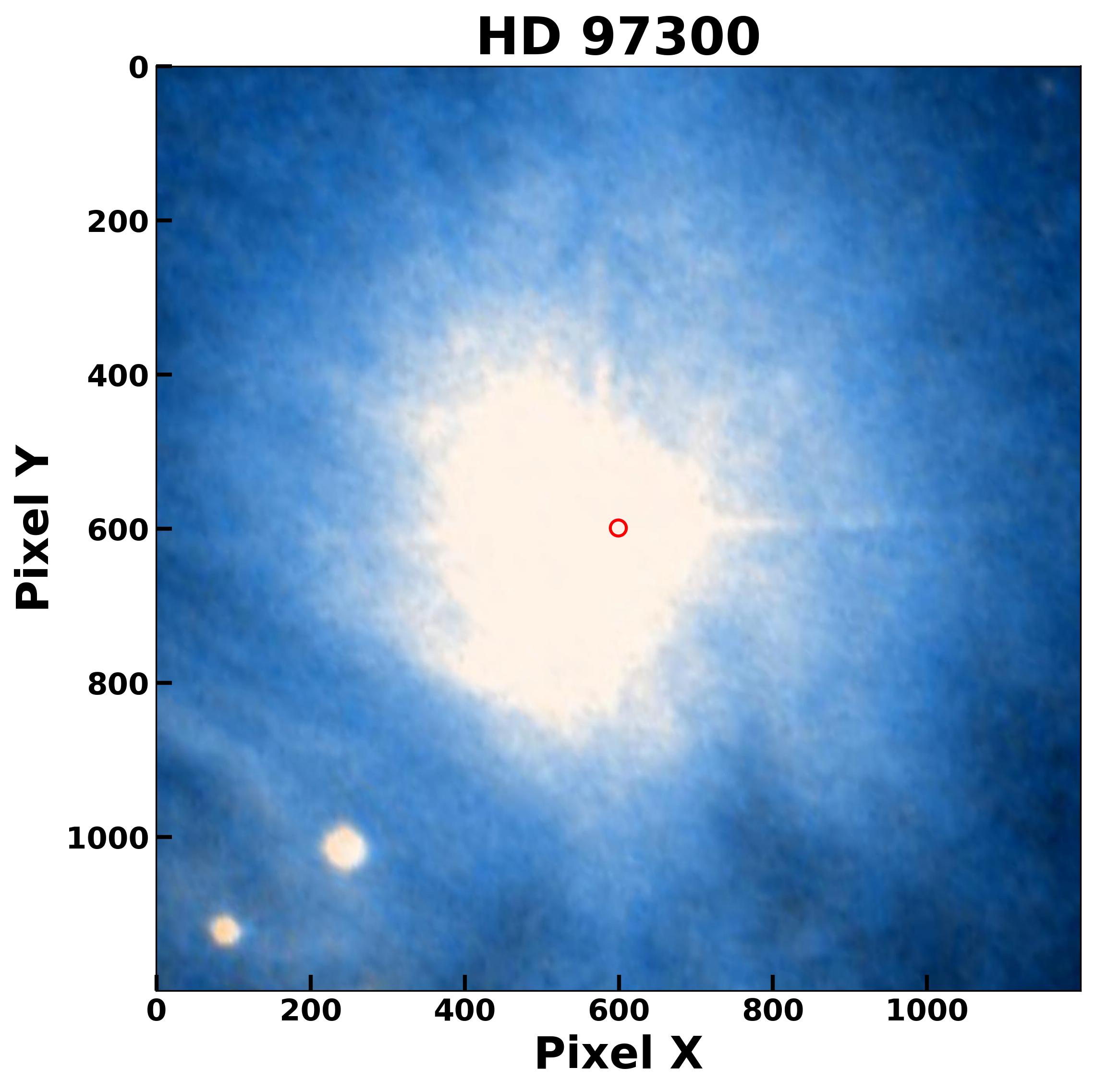}
    \includegraphics[width=0.32\columnwidth]{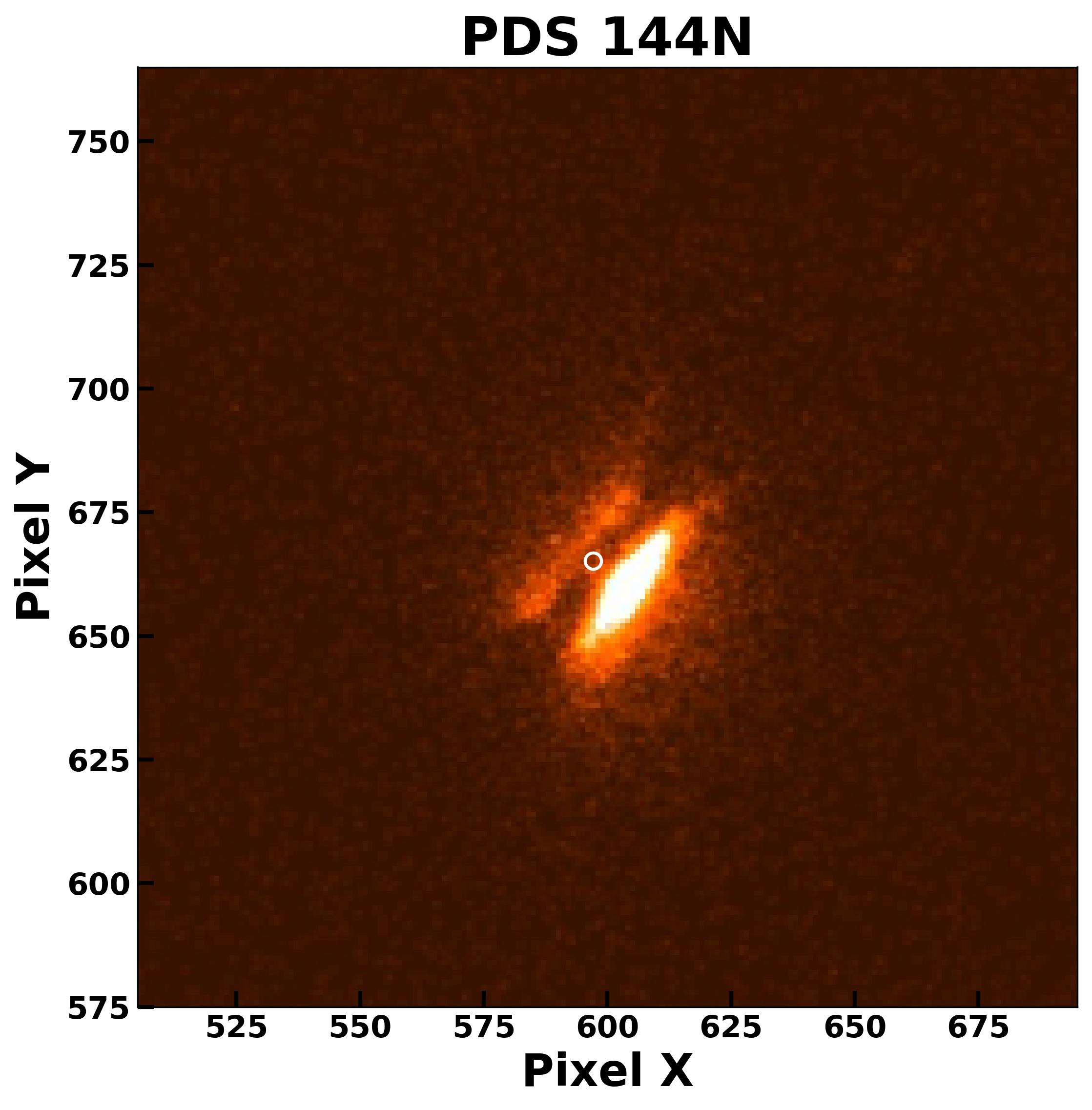}
    \includegraphics[width=0.32\columnwidth]{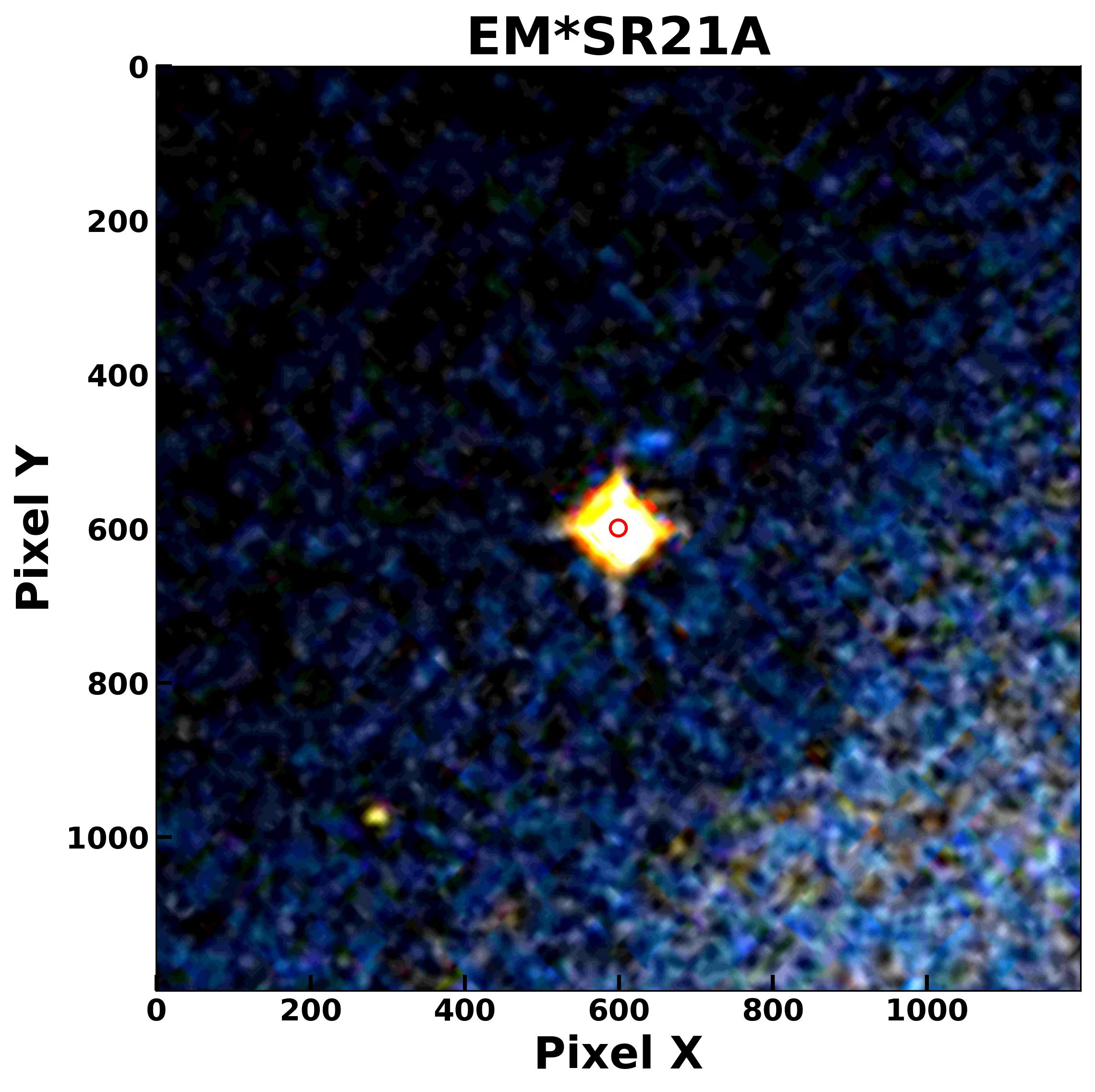}
    \includegraphics[width=0.9\columnwidth]{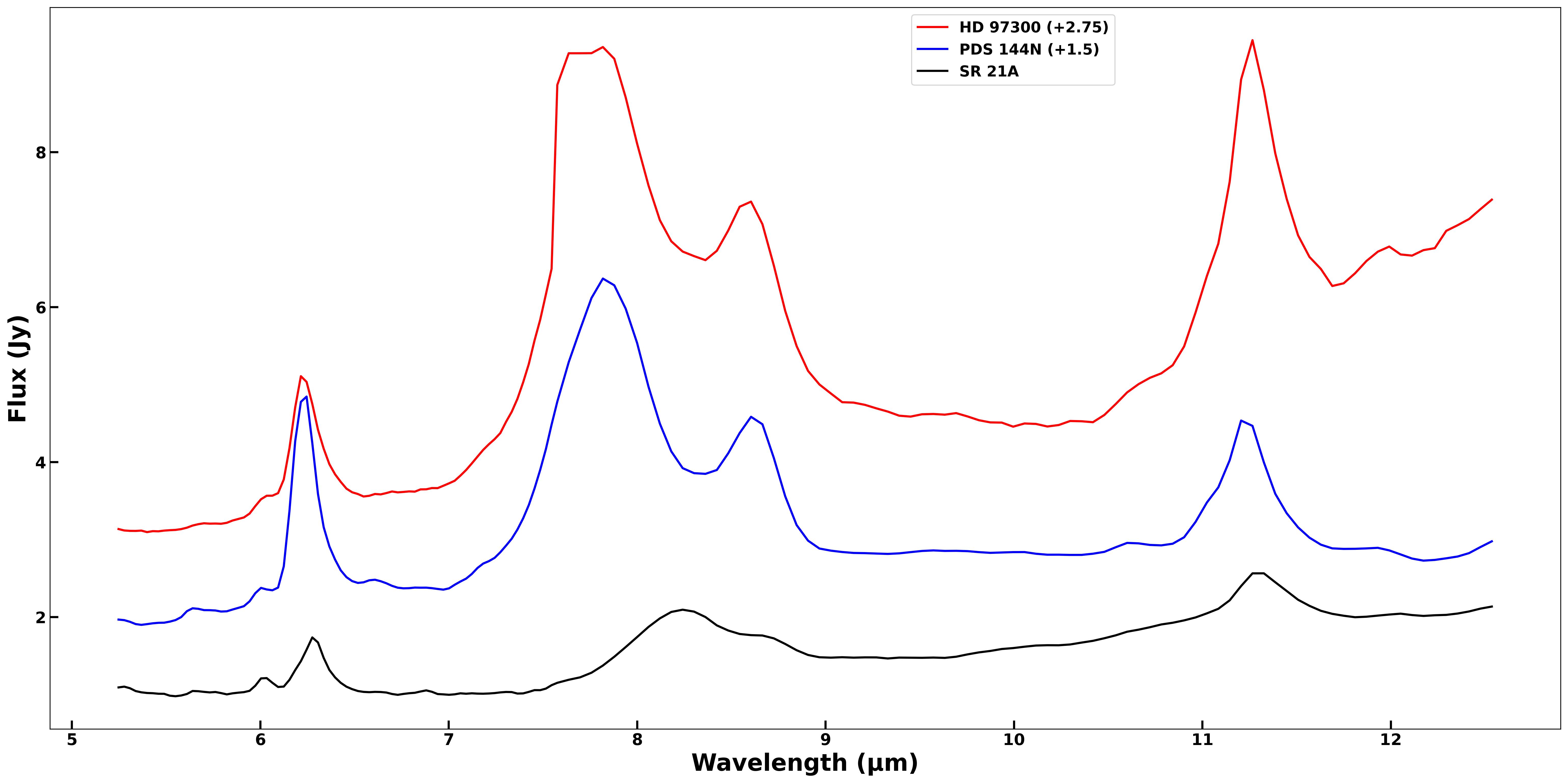}
    \includegraphics[width=0.9\columnwidth]{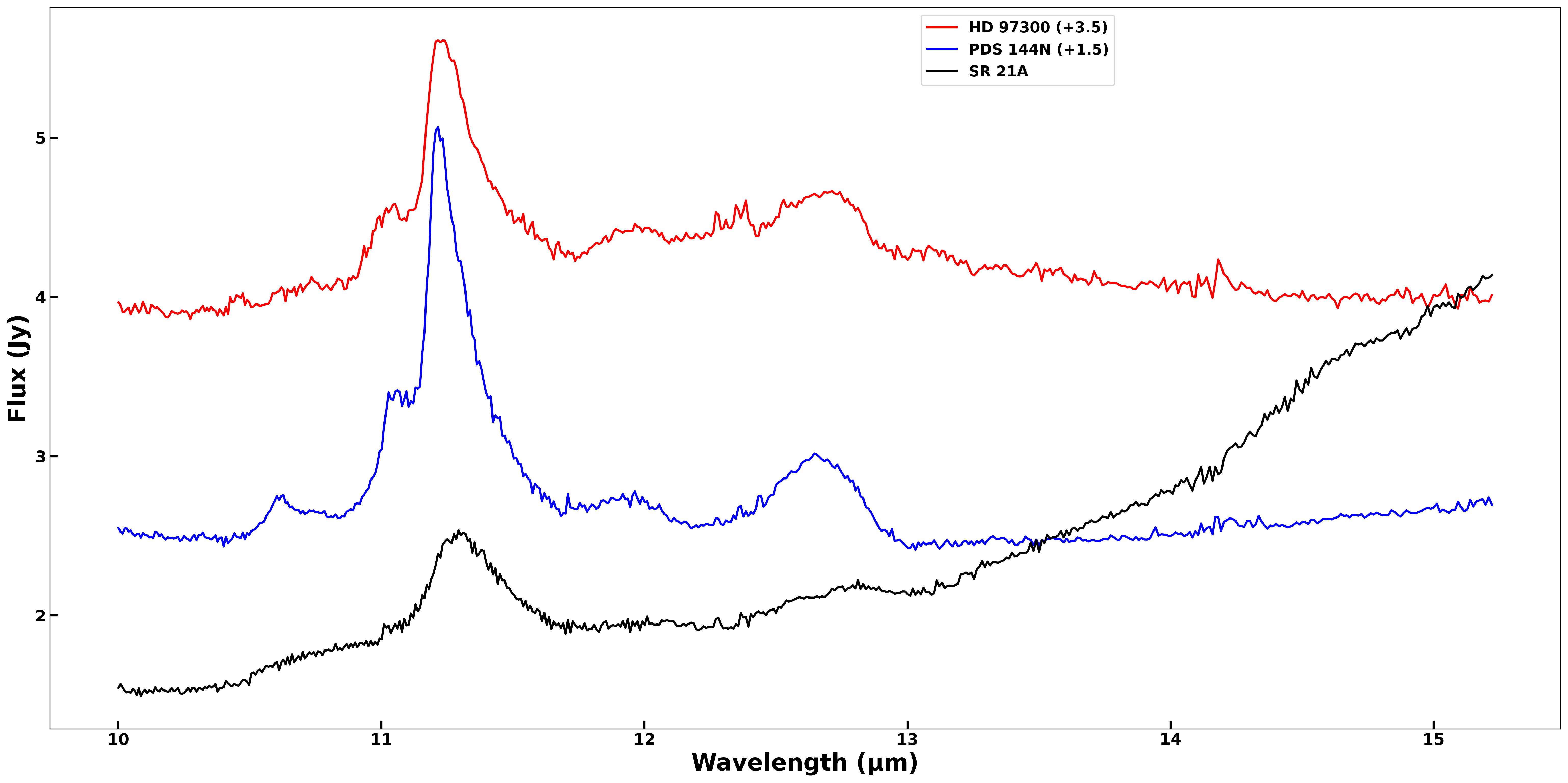}
    \caption{\textbf{Top row:} Optical/near-infrared images showcasing the morphological diversity around the three reference stars.
\emph{Left:} DSS2 color composite of \textbf{HD97300} (B9 star), highlighting its bright reflection nebulosity.
\emph{Center:} \emph{HST}/ACS F165LP image of \textbf{PDS144N} (A2 star) seen nearly edge-on, revealing a flared, dark disk silhouette.
\emph{Right:} DSS2 color composite of \textbf{SR 21A} (G1 star), an intermediate-mass T Tauri. \textbf{Middle and bottom rows:} \textit{Spitzer}/IRS spectra for the same three stars, illustrating both the \emph{low-resolution} (SL/LL) and \emph{high-resolution} (SH/LH) modules. Each spectrum demonstrates prominent mid-infrared PAHs features (e.g., at 6.2, 7.7, 8.6, 11.2,$\mu$m).}
    \label{fig:reference}
\end{figure}

This trend is consistent with the understanding that higher-energy photons emitted by hot, early-type HAeBe stars, particularly those with effective temperatures exceeding 15000 K, are likely to disrupt or modify the structure of circumstellar silicate grains. And the results align with earlier studies on silicate emission in HAeBe stars. Studies by \cite{Bouwman2001A&A...375..950B} and \cite{Meeus2001A&A...365..476M} found that silicate features are more prominent in stars of intermediate masses (like Herbig Ae stars) but tend to weaken or disappear in B-type Herbig stars. Silicate grains, which are common constituents of protoplanetary disks, tend to be smaller and more fragile in environments with strong UV radiation. As the temperature increases, the extreme UV flux associated with these stars is thought to evaporate or alter the grain structure, leading to their crystallization or total destruction, reducing the prevalence of observable silicate features \citep{Meeus2001A&A...365..476M, Bouwman2001A&A...375..950B, vanBoekel2005A&A...437..189V}. Interestingly, T Tauri stars, which are cooler and less luminous than HAeBe stars, often show strong silicate emission features. For instance, studies like \citet{Furlan2006ApJS..165..568F} and \citet{Kessler2006ApJ...639..275K} reported higher silicate detection rates in T Tauri stars (80\%), attributed to the lower intensity of UV fluxes around these stars.

When compared with PAH detection around HAeBe stars, the trends in silicate detection across T\textsubscript{eff} are somewhat similar. Both PAHs and silicates tend to show higher detection rates in cooler stars (A-type stars), and both features are suppressed in hotter B-type stars. The primary difference lies in the fact that PAHs are more susceptible to complete destruction by UV radiation, whereas silicates tend to become crystallized or modified rather than completely destroyed.
\subsection{Reference Stars with Strong PAH Emission} \label{sec:refstars}

In addition to our main sample of HAeBe stars from the SSHC, we selected three reference objects to serve as comparative benchmarks for different PAH-exciting environments. These sources were chosen because they each possess both high- and low-resolution \textit{Spitzer}/IRS spectra in the CASSISjuice archive, allowing a more detailed view of the mid-IR emission features and providing a check on our classification and fitting methods. Specifically, these three stars exemplify a range of spectral types, and disk geometries in which PAHs can be excited:

\begin{itemize} \item \textbf{HD 97300 (Herbig Be star).}
This B9 spectral type, intermediate-mass star, located in Chamaeleon I star-forming region is embedded in a reflection nebula (RNe) that shows intense PAH features in the mid-infrared \citep{Keller2008ApJ...684..411K,Manoj2011ApJS..193...11M}. HD 97300 shows a
class A PAH spectrum due to its association with RNe. Additionally, the detection of fullerenes (C\textsubscript{60}) around HD 97300 further highlights the complexity of the organic chemistry in star-forming regions \citep{Roberts2012MNRAS.421.3277R,Arun2023MNRAS.523.1601A}.

\item \textbf{PDS 144\,N (Herbig Ae star).}
The PDS 144 system comprises two pre-main sequence stars, PDS 144\,N and PDS 144\,S, which form a well-known binary separated by 5 arcseconds on the sky and at a distance of   145 $\pm$ 2 pc, are members of the Upper Sco association \citep{Hornbeck2012ApJ...744...54H}. The northern component, PDS 144\,N, is viewed nearly edge-on, revealing a highly flared disk geometry \citep{Perrin2006ApJ...645.1272P}, with a spectral type of A2. Prominent PAH emission is detected in PDS 144N using imaging and TIMMI2 N-band spectra were obtained between 8–13 $\mu$m \citep{Perrin2006ApJ...645.1272P,Schutz2009A&A...507..261S}. The \textit{Spitzer} IRS spectra of the star is not previously used in any studies.  The simultaneous availability of SL/SH and LL/LH data for PDS 144\,N enables refined comparisons of line-to-continuum ratios and PAH band profiles, making it an excellent case study for investigating how flared disk structures enhance PAH excitation.

\item \textbf{SR 21A (intermediate-mass T~Tauri).} SR 21A, identified as a G2.5-type T Tauri star located in the Ophiuchus star-forming region, situated approximately 125 pc away. SR 21A is one of the few T Tauri stars that show clear 11.2 $\mu$m PAH features \citep{Geers2006A&A...459..545G}. The low resolution spectra of SR 21A is not studied in previous studies. The spectra shows all the PAH features. The star is characterized by having "cold disks" with SED that lack significant excess emission in the 3–13 $\mu$m region, indicating the presence of an inner hole in their dust disks. The presence of this gap enhances the feature-to-continuum ratio of PAH emissions, making them more detectable. The presence of all PAH features makes the spectra a good reference for a low mass disk with PAH, for comparison.  
\end{itemize}

\noindent The spectra of these three reference stars in the \autoref{fig:reference}  illustrating both the low-resolution SL/LL data and the high-resolution SH/LH modules. In addition, images of three stars that reveal the morphology of their surrounding nebulosity or disk structure. The DSS2/color images of HD 97300 and SR21A and HST/ACS/F165LP of PSD 144N is also shown in \autoref{fig:reference}.  Collectively, these three objects serve as comparative benchmark for PAH emission in diverse pre-main sequence environments, relative to the main HAeBe sample. 

\subsection{6.2 $\mu m$ PAH Feature}
Building on the earlier analysis of spectral classification of our SSHC catalog, we now focus on the 64\% of the sources exhibiting PAH feature in their spectra. The most common PAH feature detected is the 6.2 $\mu m$ band, which arises from C-C stretching vibrations in PAHs \citep{Peeters2002A&A...390.1089P,Smith2007ApJ...656..770S}, observable with the \textit{Spitzer} IRS in low-resolution mode. Of our sample, 50 sources exhibited the 6.2 $\mu m$ band, the largest number for any single feature. A weaker feature at 6.0 $\mu m$, blended with the primary 6.2 $\mu m$ peak, was also identified. This weaker feature has traditionally been attributed to C=O (carbonyl) stretching vibrations \citep{Peeters2002A&A...390.1089P}, though its origin remains debated. Recent studies have proposed that it may instead arise from olefinic C=C bonds \citep{Hsia2016ApJ...832..213H}. We detected the 6.0 $\mu m$ band in 38 of the 50 sources with the 6.2 $\mu m$ feature.

Numerous studies have probed shifts in the peak positions of PAH features as a function of  T\textsubscript{eff} of the central star. As established by \cite{Acke2010}, the peak wavelengths of PAH features in Herbig Ae stars show variations correlated with T\textsubscript{eff}. Leveraging our large sample, we examined the 6.2 $\mu m$ feature shift with T\textsubscript{eff} and incorporated higher-mass Herbig Be stars to explore whether PAH peak shift trends are consistent across both Herbig Ae and Be objects. Examining higher-mass stars is essential, as factors like disk evolution, accretion mechanisms, magnetic fields, UV output, and disk structure differ substantially between early and late-type pre-main sequence stars \citep{Mottram2007,Alonso2009A&A...497..117A,Fairlamb2017MNRAS.464.4721F,Shridharan2023JApA...44...62S}, and the PAH processing can get affected due to these factors. Thus, potential variations in PAH emissions between Herbig Ae and Be stars warrant investigation.

We extracted the 5-7 $\mu m$ region of the low resolution IRS spectra for 50 HAeBe stars with 6.2 $\mu m$ features. To fit the spectra and get the peak wavelength and flux, a continuum subtraction is needed. Using cubic spine to define a continuum is adopted in various studies on PAH emission from ISO, \textit{Spitzer} and JWST \citep{vanDiedenhoven2004ApJ...611..928V,Seok2017ApJ...835..291S,Zhang2024arXiv241018909Z}. We adopted a cubic spline function to subtract the continuum over this wavelength range. For 38 stars, we fit the 6.2 and 6.0 $\mu m$ features with a double Gaussian model, deblending the peaks. The 6.2 $\mu m$ band exhibits asymmetry in its blue wing, hence a simple gaussian or drude profile cannot accurately recover the integrated flux. By fitting dual gaussians and subtracting the fitted 6.0 $\mu m$ flux from the total integrated flux between 5.85 $-$ 6.8 $\mu m$ region, we isolated the 6.2 $\mu m$ flux \citep{Peeters2017ApJ...836..198P}. For sources without a 6.2 $\mu m$ detection, a single Gaussian fit to get peak wavelength and integration for flux was implemented. Figure. \ref{fig:fit} in appendix illustrates example dual and single fits to the 6.2 $\mu m$ PAH complex. In addition to the sample, we estimated the 6.2 $\mu m$ parameters for the three reference stars.

\begin{figure}
    \centering
    \includegraphics[width=0.49\columnwidth]{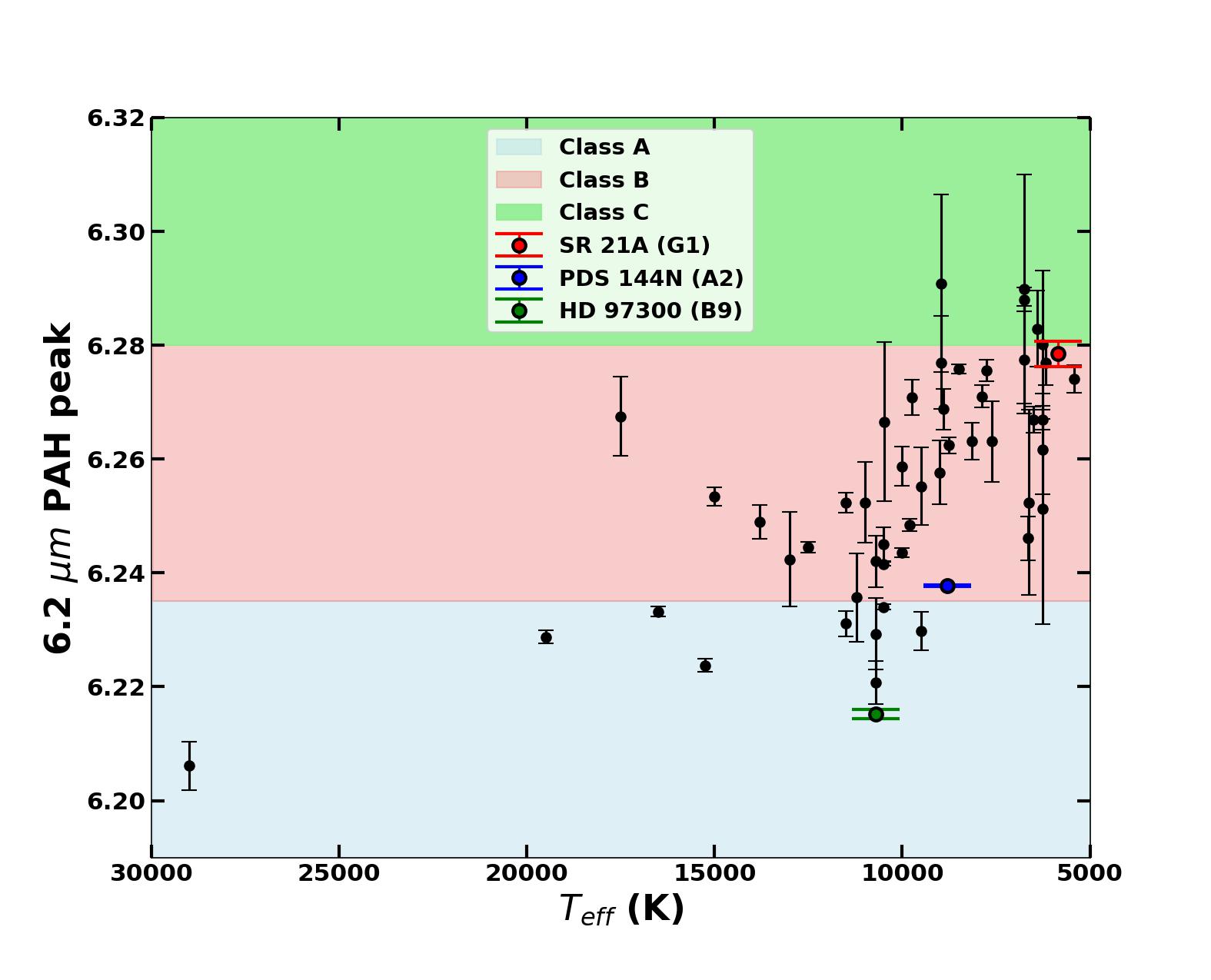}
    \includegraphics[width=0.49\columnwidth]{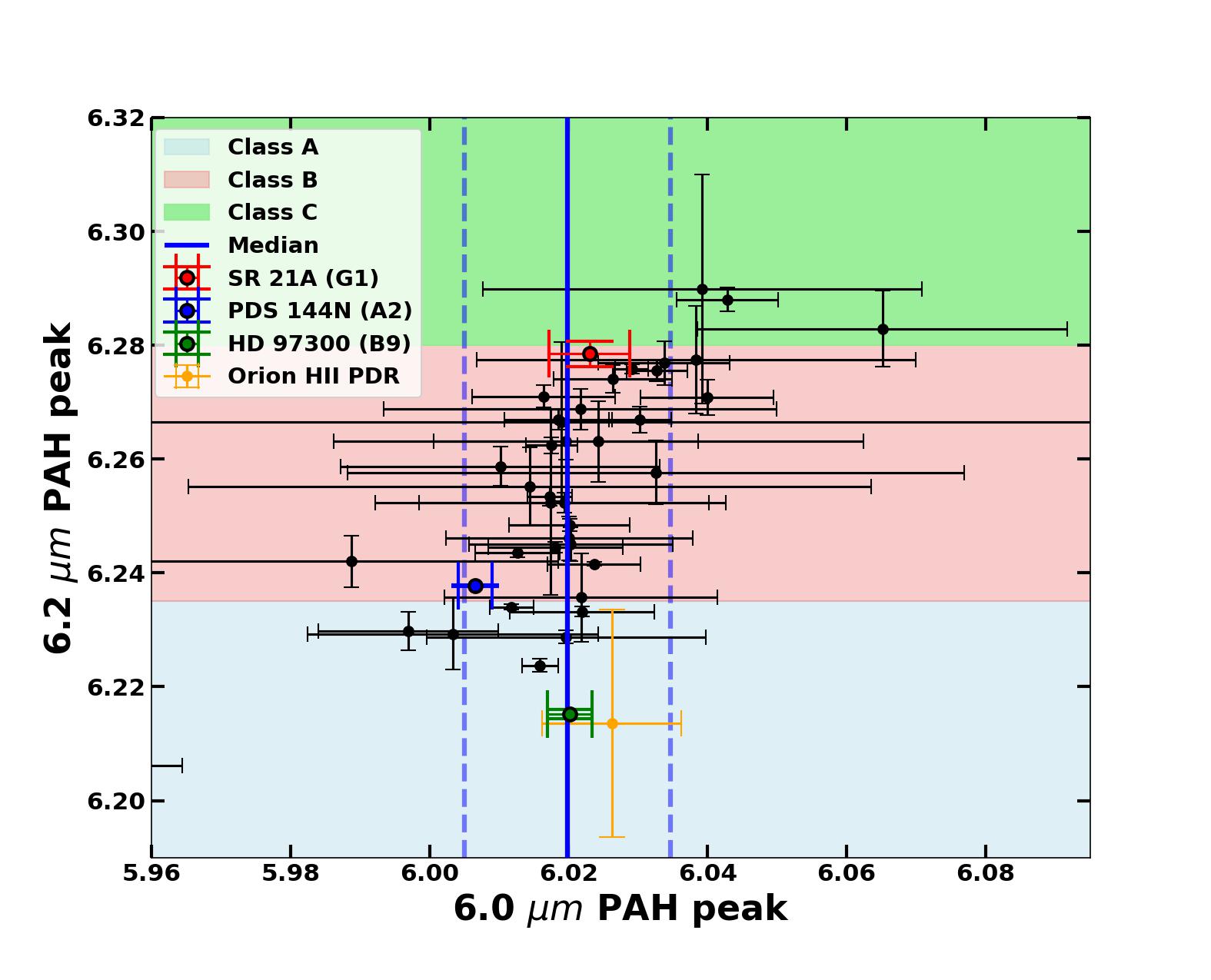}
   \caption{Left panel: The peak wavelength of the 6.2~$\mu$m PAH feature shifts with the effective temperature ($T_{\text{eff}}$) of the central star, indicating a transition from Class B/C (circumstellar) to Class A (nebular) profiles. Right panel: The 6.0~$\mu$m feature, attributed to C=O stretching, remains nearly invariant (median $\approx$6.02~$\mu$m) across the sample, consistent with Orion PDR \textit{JWST} template spectra.}
    \label{fig:pah6}
\end{figure}

Figure \ref{fig:pah6} presents the distribution of the peak wavelengths for the 6.2 $\mu$m PAH feature in our HAeBe sample. The peak wavelength of the 6.2 $\mu$m PAH shifts to longer values as the $T_{\text{eff}}$ of the central star decreases. A similar conclusion was drawn by \cite{Acke2010} for 7.7 $\mu$m PAHs for a sample of HAeBe stars. While the 6.2 and 7.7 $\mu$m PAH features often show correlated shifts across different environments, the physical causes of these redshifts are not identical. The 6.2 $\mu$m band originates from aromatic C–C stretching vibrations and its redshift is primarily attributed to chemical modifications of the PAH molecules—such as the incorporation of heteroatoms (e.g., nitrogen), partial hydrogenation, or the presence of aliphatic sidegroups \citep{Peeters2002A&A...390.1089P, Pino2008A&A...490..665P}. In contrast, the 7.7 $\mu$m complex—although it often shifts alongside the 6.2 $\mu$m band—comprises a blend of C–C and C–H in-plane modes whose position is strongly influenced by the size distribution and ionization state of the emitting PAH population \citep{Bauschlicher2008ApJ...678..316B, Maragkoudakis2023MNRAS.520.5354M}. With 50 stars exhibiting the 6.2 $\mu$m band, this is the largest sample tested for this correlation. PAH features are categorized into Class A and Class B based on their peak wavelength positions and profiles.

Class A PAH bands peak between 6.19 and 6.235 $\mu$m and are predominantly found in the ISM or planetary nebulae. Class B PAH bands have profiles with a peak position between 6.235 and 6.28 $\mu$m and are typically found in Herbig Ae disks and class C spectra which is highly redshifted (6.28 -- 6.32) is seen in low mass T Tauri disk \citep{Peeters2002A&A...390.1089P}. These classifications are illustrated in Figure \ref{fig:pah6} along with the three reference stars. The two stars PDS 144N and SR21A shows class B PAHs, denoting that the PAHs are originated from circumstellar region. The HBe star HD 97300 shows class A type PAH as it has associated nebulosity \citep{Keller2008ApJ...684..411K}.

In Herbig Ae stars, we observe a gradual transition from Class B to Class A PAHs as $T_{\text{eff}}$ increases, indicative of changes in the molecular composition of the PAH population with increasing radiation field. This trend likely reflects the progressive chemical processing of PAHs under stronger UV radiation: ultraviolet photons can destroy aliphatic sidechains, dehydrogenate PAHs, or fragment smaller and less stable molecules, leaving behind more compact, stable aromatic species. Such UV-driven evolution results in PAH spectra shifting toward Class A profiles, characterized by shorter-wavelength features.

\citet{Maragkoudakis2023MNRAS.520.5354M} demonstrated through DFT modeling that the 7.7 $\mu$m complex shifts to longer wavelengths as the average PAH size increases. However, in stronger UV environments, smaller PAHs are more likely to be photo-destroyed, resulting in a surviving population of more stable, medium-sized ionized PAHs that dominate the observed emission at ~7.6 $\mu$m (Class A). Conversely, in lower UV environments, larger and more neutral PAHs persist, producing the redshifted 7.8–8.0 $\mu$m Class B/C profiles.

The sample of Herbig Be stars, however, exhibit a combination of both Class A and Class B PAH features within the sample, irrespective of the star’s $T_{\text{eff}}$. This simultaneous presence of both PAH classes in Herbig Be stars suggests additional influences beyond the star’s UV output. The \textit{Spitzer} IRS spectra integrate the complete circumstellar region around the source, and many higher-mass Herbig Be stars in our sample exhibit associated nebulosity along with their disks \citep{Arun2023MNRAS.523.1601A}. Specifically, contributions from surrounding reflection nebulae or interstellar PAH sources may account for the observed variety in PAH profiles. Reflection nebulae can introduce Class A features depending on the nebular conditions, while the disk may contribute Class B-type spectra. Therefore, in the case of Herbig Be stars, the observed PAH profiles likely represent a mixture of direct circumstellar and nebular contributions, each affecting the PAH population differently. Spatially resolved studies are important in disentangling the nebular and disk PAHs and their nature.

The 6.0 $\mu$m weak feature was proposed to not correlate with the 6.2 $\mu$m peak shift due to its origin from C=O (carbonyl) or olefinic C=C stretching vibrations. We find that the 6.0 $\mu$m feature does not shift with the 6.2 $\mu$m feature. Figure \ref{fig:pah6} (right) shows the peak wavelength of the 6.2 and 6.0 $\mu$m features of 38 HAeBe stars in our sample. The figure shows that most of the data points are within 3 times the median absolute deviation (MAD) of the sample. The median value of the 6.0 $\mu$m feature is 6.02, which, along with MAD, is denoted in the region. The HII region template JWST spectra \citep{Chown2024A&A...685A..75C} obtained from the Orion PDR is fitted to find the 6.0 $\mu$m feature and the peak wavelength of the 6.2 $\mu$m feature; both of them are shown in the figure. The Orion PDR has Class A type PAHs, and the 6.0 $\mu$m feature peaks at 6.026 $\pm$ 0.02 $\mu$m, consistent with our data.

Further studies are required to investigate the chemical diversity and environmental conditions that contribute to the presence and strength of the 6.0 $\mu$m feature. Detailed modeling and high-resolution spectroscopy across various astrophysical environments will be crucial to fully understand the role of C=O functional groups in producing this weak feature.

\subsection{PAH Band Ratios}

The relative intensities of key PAH bands, expressed as ratios such as 6.2/11.2, 7.7/11.2, 8.6/11.2, and $12.7/11.2$\ reveal insights into the ionization state and size distribution \citep{Tielens2008ARA&A..46..289T,Keller2008ApJ...684..411K}. We selected 21 stars from our catalog with classifications of PAH-only (P) and PAH+Silicate (PS), focusing on sources that exhibit prominent 7.7, 8.6, and 11.2 $\mu$m PAH features well above the continuum. The continuum was subtracted using spline anchor points as defined in \citet{Seok2017ApJ...835..291S}. The integration of PAH features was performed over the wavelength ranges specified in \citet{Zhang2024arXiv241018909Z}, who studied PAH emission in the 30 Doradus region using JWST observations. Following this, we calculated the ratios of 6.2/11.2, 7.7/11.2, 8.6/11.2, and $12.7/11.2$\  PAH bands to investigate their relative intensities and correlations, offering insights into the physical and chemical properties of PAHs.

\begin{figure}
    \centering
    \includegraphics[width=0.9\columnwidth]{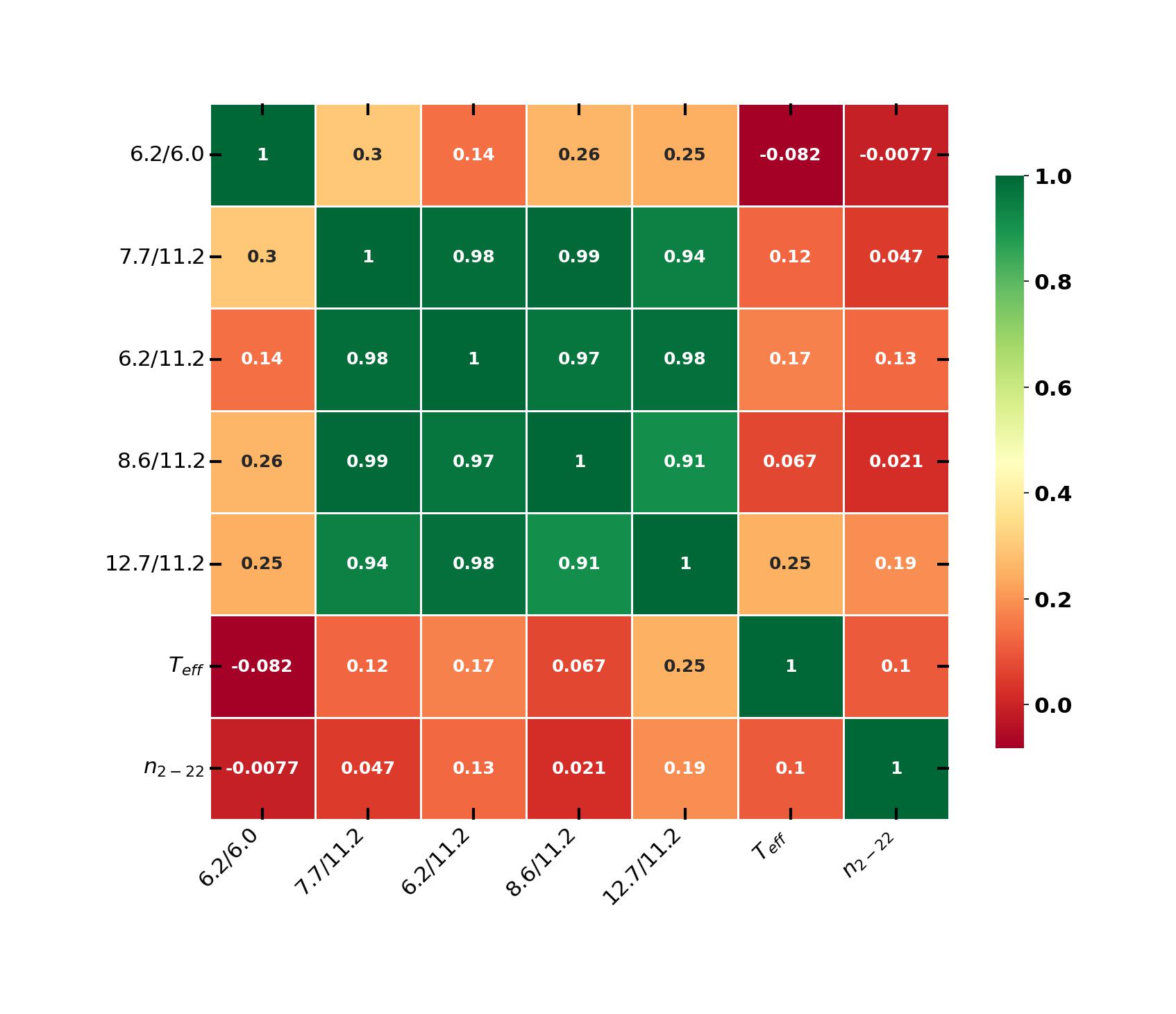}

    \caption{Correlation matrix of selected PAH band ratios $T_{\text{eff}}$, and $n_{2-24}$ for HAeBe stars. The heatmap shows the Pearson correlation coefficients between the parameters. The strong positive correlations ($r \sim 0.9 - 0.98$) between $6.2/11.2$, $7.7/11.2$, $8.6/11.2$, and $12.7/11.2$ suggest that these band ratios co-vary due to shared dependence on the size distribution and molecular structure of the emitting PAHs. While PAH ionization influences these features, the ratios are more strongly governed by PAH size and the number of peripheral C–H bonds. The color intensity represents the strength of the correlation, with red indicating negative correlations and green indicating positive correlations.}

    \label{fig:correlation_heatmap}
\end{figure}

We computed a correlation matrix, presented in Figure \ref{fig:correlation_heatmap}. This matrix includes the PAH band ratios $6.2/6.0$, $6.2/11.2$, $7.7/11.2$, $8.6/11.2$, $12.7/11.2$, $T_{\text{eff}}$ and, $n_{2-24}$. The Pearson correlation coefficient was used to quantify the degree of linear relationship between these variables.

The heatmap reveals distinct trends and correlations among the PAH ratios and the effective temperature:

\begin{itemize}
    \item The $6.2/11.2$, $7.7/11.2$, $8.6/11.2$, and $12.7/11.2$\ ratios show strong positive correlations with each other ($r \sim 0.9 - 0.99$). 

    \item The $6.2/6.0$ ratio exhibits weaker correlations with the other PAH ratios and $T_{\text{eff}}$. This suggests that this ratio, originating from carbonyl or olefinic C=C stretching stretching vibration is influenced by different processes compared to the other bands.

\end{itemize}

The heatmap analysis highlights the importance of the local UV environment in shaping PAH emission. The strong correlations among the $6.2/11.2$, $7.7/11.2$, $8.6/11.2$, and $12.7/11.2$ ratios indicate that the relative strengths of these PAH features are influenced by shared physical factors, mainly the size and molecular structure of the PAH population, which evolve under UV processing. Although PAH ionization state does affect these bands, especially the 11.2 $\mu m$ C–H out-of-plane mode relative to 7.7 $\mu m$ C-H in-plane bending modes \citep{Maragkoudakis2020MNRAS.494..642M,Ujjwal2024A&A...684A..71U}, the dominant drivers of these ratios are often changes in PAH size and edge structure. Meanwhile, the weak correlations between $T_{\text{eff}}$ and PAH ratios imply that PAH emission is more responsive to the radiation field in the immediate surroundings and disk structure than the stellar properties themselves. These findings are in agreement with earlier studies \citep{Tielens2008ARA&A..46..289T,Sloan2005ApJ...632..956S}.

\subsection{11.0/11.2 $\mu$m PAH Band Ratio}

The 11.0 $\mu$m and 11.2 $\mu$m features of PAHs are produced by CH out-of-plane bending modes \citep{Perrin2006ApJ...645.1272P,Candian2015MNRAS.448.2960C}. The weak feature at 11.0 $\mu$m has been attributed to the CH out-of-plane bending mode of lone hydrogen groups in PAH cations\citep{Hony2001A&A...370.1030H}. Thus, the 11.0/11.2 $\mu$m band ratio are widely recognized as valuable tracers of PAH ionization states in various astrophysical environments \citep[e.g.,][]{Tielens2008ARA&A..46..289T}. In SSHC, we have mid-infrared spectra of 76 HAeBe stars with the high-resolution module of the \textit{Spitzer}, 53 of which exhibit detectable 11.2 $\mu$m PAH emission. Among these, 20 sources display resolved 11.0 $\mu$m and 11.2 $\mu$m PAH bands, allowing for a robust investigation of the ionization state of PAHs in these stellar environments. The strong silicate feature obscure the 11.0 $\mu$m feature, complicating its detection in other cases.

\begin{figure}
    \centering
    \includegraphics[width=0.75\columnwidth]{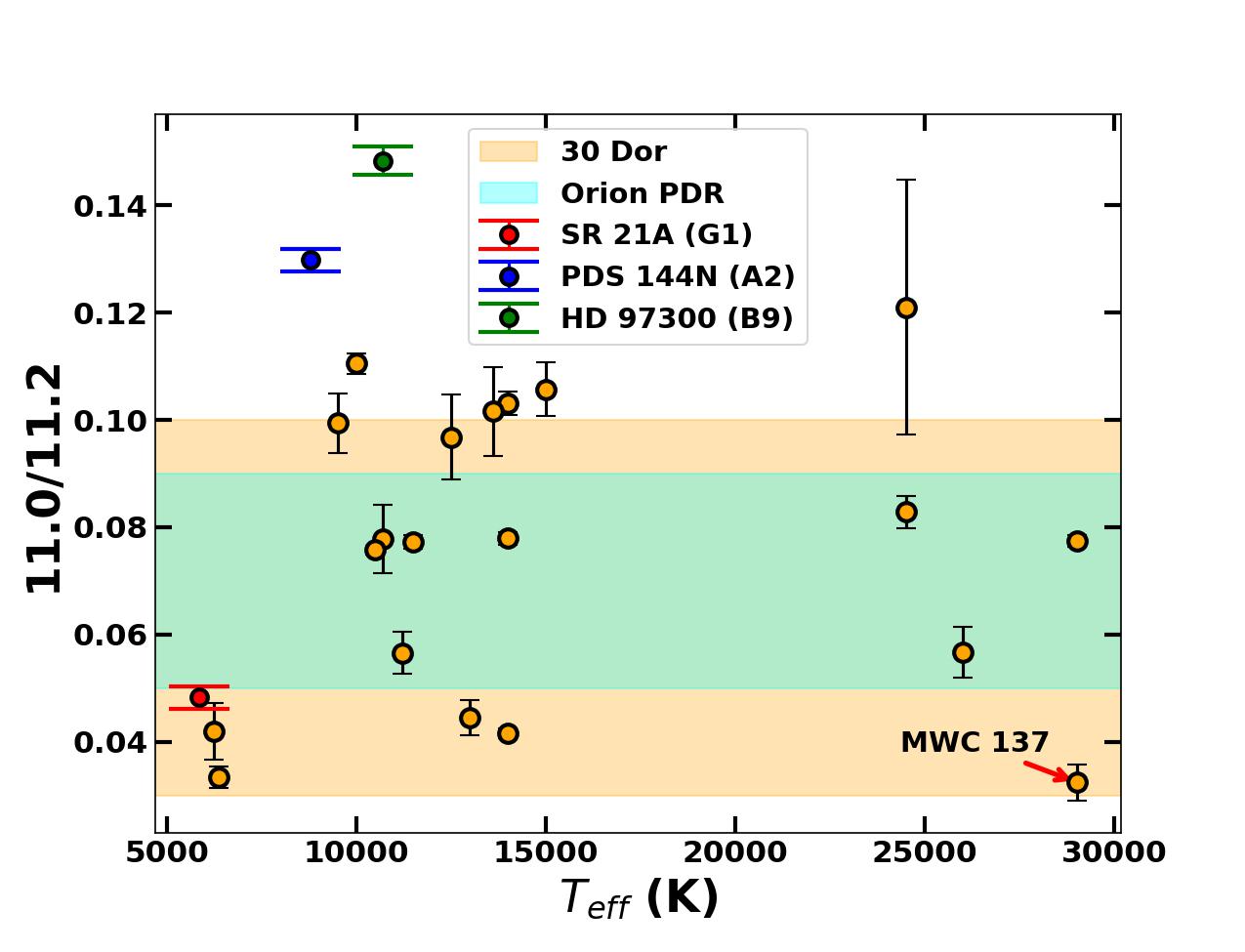}
    \caption{Distribution of the 11.0/11.2 PAH band ratio versus stellar effective temperature for our HAeBe sample. Reference stars (SR 21A, PDS 144N, HD 97300) shows different ionization regimes. The shaded regions approximate PDR‐like ranges from JWST data, whereas very low ratios occur at 5000–6000 K suggestive of neutral PAHs in low mass disk. MWC 137, which is showing very low ionization is likely an evolved B[e] star rather than an HAeBe star. }    
    \label{fig:ion}
\end{figure}

We continuum-subtracted the spectra and calculated the integrated flux of the 11.0 $\mu$m and 11.2 $\mu$m bands using defined wavelength ranges of 10.9–11.13 $\mu$m and 11.13–11.75 $\mu$m, respectively (similar approach as \citealp{Zhang2024arXiv241018909Z}). We used 10.0, 10.5, 10.87, 11.7, 13.0, 14.0, 15.20 $\mu$m as spilne anchor points for the subtraction. An example of the flux extraction is given in Appendix (\autoref{fig:11.2_sub}). To our knowledge, this study represents the first systematic analysis of this ratio in a sample of HAeBe stars. The resulting 11.0/11.2 $\mu$m flux ratios span a range of 0.03–0.12. This variability indicates differences in the ionization conditions across the observed sample of HAeBe stars. \autoref{fig:ion} shows 11.0/11.2 PAH ratio with T\textsubscript{eff} of the star.

When compared to the recent observations of PDRs in Orion and 30 Dor (LMC), which exhibits an 11.0/11.2 $\mu$m ratio of in the range of 0.05$-$0.9 and 0.03$-$0.1 respectively \citep{Chown2024A&A...685A..75C, Zhang2024arXiv241018909Z}. The ratio of HAeBe stars demonstrate that several sources fall within a similar ionization regime as PDRs, while others show higher ionization fractions. Interestingly, HD 97300 and PDS 144N are shows highest ionization (0.16 and 0.14) among the list of stars in the study, and SR21A shows one of the lowest ionization. PDS 144N is showing highest ionization among circumstellar disk PAHs, most likely due to the flared morphology enhancing the UV irradiation. Thus, the ionization of PAHs in PDRs and circumstellar disks are of similar values, but some sources especially HAe disk show higher fraction of ionization.

At the high end of the observed ratio range (e.g.\ 0.16), ionized PAHs appear to dominate under stronger UV fields, as in A-type stars or B-type stars with circumstellar reflection nebulae. Conversely, lower ratios (e.g.\ 0.03) suggest predominantly neutral PAHs, as seen in low-mass disks such as SR21A. In conjunction with results from Sec.\ 3.5, these findings indicate that circumstellar disks around Herbig Ae/Be stars generally host more processed and ionized PAHs than those found in nebulae or the ISM. Notably, three stars with T\textsubscript{eff} $\simeq$ 5000–6000 K exhibit very low 11.0/11.2 ratios, implying largely neutral PAHs in these lower-mass disks. The broad spread of ratios in Herbig Ae/Be disks likely reflects variations in proximity and disk geometry. Interestingly, MWC137 shows a low ratio despite its high temperature; although often catalogued as an HAeBe star, it has been classified as an evolved B[e] object \citep{Muratore2015AJ....149...13M,Mehner2016A&A...585A..81M}, and its expanding shell may still contain newly formed, un-ionized PAHs. Hence, MWC137 should be excluded from future HAeBe analyses.

The circumstellar disks have stronger radiation field due to the proximity of PAHs to the ionizing source. The disks are denser than the ISM, and some having flared morphology facilitating interactions that favor ionization. The processing of PAHs through photodissociation and ionization is efficient in circumstellar disks, altering their charge states significantly. Spatial information on the circumstellar PAHs around HAeBe star will help in understanding the transformation and evolution of PAHs in circumstellar disk and envelopes.

\section{Conclusion}

In this work, we have presented a comprehensive mid-infrared analysis of HAeBe stars using the \textit{Spitzer} IRS spectral catalogue. We classified 124 stars based on their molecular spectral features, categorizing them into PAH-only, silicate-only, PAH and silicate, HACs, and featureless spectra. Our classification reveals a diversity of circumstellar environments, with PAHs and silicates playing distinct roles in the disks surrounding these stars.

We found that HAeBe stars with a spectral index $n_{2-24} > -1$ exhibit a higher likelihood of showing PAH features. Also, the Meeus group I sources are identified to have higher fraction of PAH detection, which is consistent with literature. Our study also highlighted an effective temperature range between 7000 K and 11,000 K where PAH detection frequencies are highest, suggesting an optimal environment for PAH survival and excitation. This detection frequency declines at higher temperatures, which could be attributed to increased UV radiation leading to the destruction or processing of PAHs. A similar trend in detection is found in the case of silicate emission.

The 6.2 $\mu$m PAH is found to be the most common in our sample of low resolution spectra. The analysis of PAH feature shifts demonstrated a correlation between the 6.2 $\mu$m peak position and the stellar effective temperature, supporting the idea that smaller PAH molecules dominate in hotter environments. Interestingly, the 6.0 $\mu$m weak feature did not show a correlation with the 6.2 $\mu$m peak, implying a distinct nature of C=O stretching than C-C vibrations.

Our results indicate that the PAH band ratios, such as 6.2/11.2, 7.7/11.2, and 8.6/11.2, are strongly correlated, suggesting a consistent response of PAH molecules to the ionizing environment. The shift in peak of  6.2 $\mu$m PAH with the T\textsubscript{eff} and the higher ionization ratio in lower temperature stars show that circumstellar PAHs are more processed and ionized than interstellar PAHs. 

We examined the emission in the 11.0 and 11.2\,\(\mu\)m bands, which trace the CH out-of-plane bending modes in PAHs and provide a key diagnostic of their ionization state. Although the 11.0\,\(\mu\)m feature is comparatively weak and can be blended with a steep silicate continuum, we successfully measured 21 stars with resolved 11.0 and 11.2\,\(\mu\)m PAH bands. The resulting 11.0/11.2\,\(\mu\)m ratios span 0.046--0.12, signifying a diverse range of ionization conditions among our HAeBe sample. Comparison with typical PDRs suggests that certain HAeBe disks exhibit enhanced PAH ionization, potentially due to strong UV fields and local disk conditions, such as flared morphologies. At the same time, some sources have low 11.0/11.2\,\(\mu\)m ratios, indicating partial or predominantly neutral PAHs, possibly influenced by stellar parameters.

Circumstellar PAHs gives an interesting window to study highly processed PAHs than those found in ISM. Future observations with JWST will allow for a more detailed examination of the chemical diversity, spatial distribution, and evolution of PAH molecules in these complex environments.

\begin{acknowledgements}
 BM, and SSK acknowledge the financial support from CHRIST (Deemed to be University, Bangalore) through the SEED money projects (No: SMSS-2335, 11/2023 \& SMSS-2220,12/2022) and  by the SERB project (CRG/2023/005271).
\end{acknowledgements}
\newpage
\appendix 
\section{Flux estimation}
The flux estimation procedures for 6.2 and 11.2 \,\(\mu\)m  PAH features are given in the appendix
\begin{figure*}[ht!]
    \centering
    \includegraphics[width=\columnwidth]{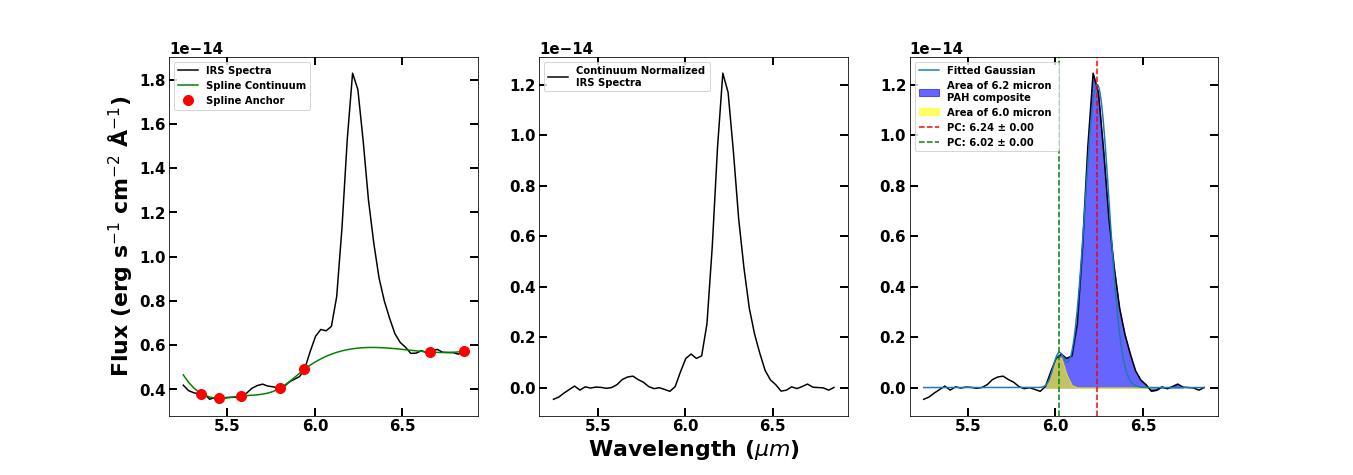}
    \includegraphics[width=\columnwidth]{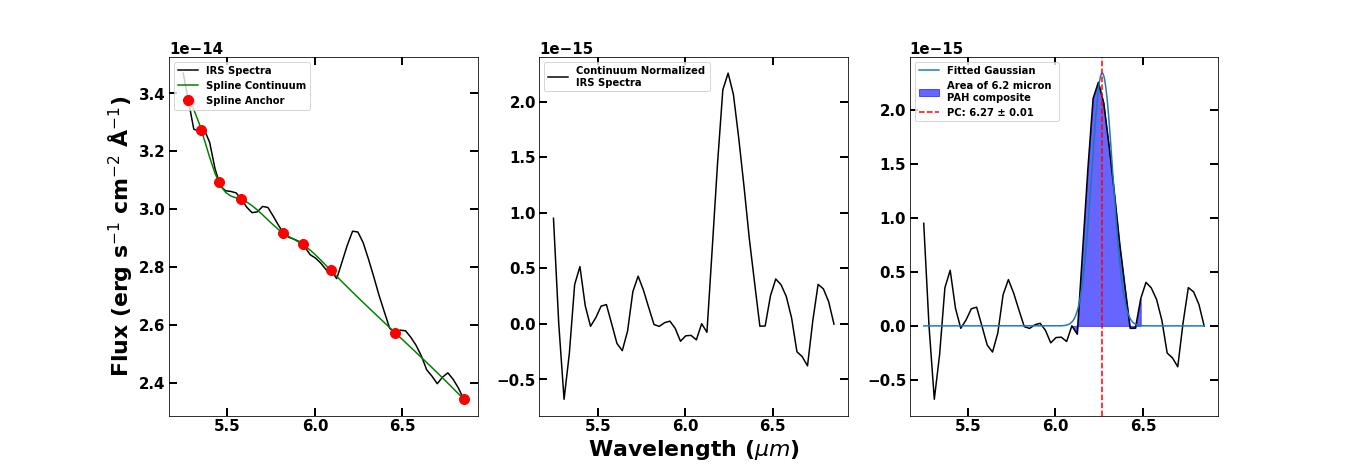}

    \caption{Representative examples of our double-Gaussian fitting approach for the 5--7\,\(\mu\)m spectral region, HD 97300 (top) and single gaussian fit PDS~37 (bottom). 
    }

    \label{fig:fit}
\end{figure*}

\begin{figure*}
    \centering
    \includegraphics[width=0.85\textwidth]{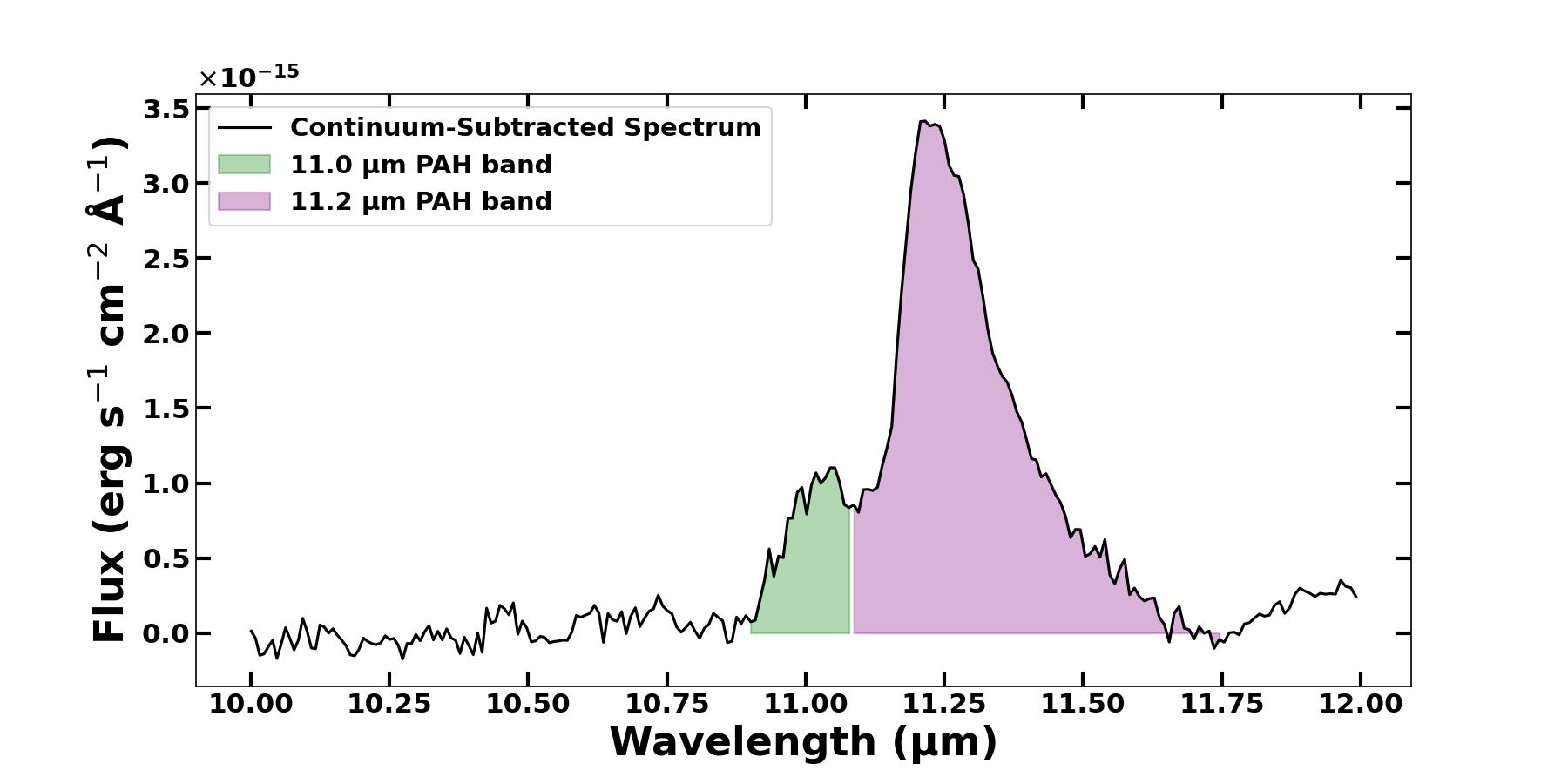}
    \caption{
    Example of the flux extraction of the 11.0 and 11.2\,\(\mu\)m PAH features in \textbf{HD~97300} Spitzer SH spectra.}
    \label{fig:11.2_sub}
\end{figure*}

\newpage

\begin{longtable}{cccccccc}
\caption{The details of 124 HAeBe stars with \textit{Spitzer} IRS spectra and their classification is given. The AOR\_LR and AOR\_HR are the \textit{Spitzer} AOR key of low and high resolution observations respectively. The new CASSISJuice ID is also given along with C\textsubscript{60} detection from \cite{Arun2023MNRAS.523.1601A}. \label{tab:table1}}\\

\hline
Source & AOR\_LR & AOR\_HR & RA (deg) & DEC (deg) & CLASS & CASSISJuice ID & C\textsubscript{60}\\
\hline
\endfirsthead

\multicolumn{7}{c}{{\bfseries Table \thetable\ continued from previous page}}\\
\hline
Source & AOR\_LR & AOR\_HR & RA (deg) & DEC (deg) & CLASS & CASSISJuice ID & C\textsubscript{60}\\
\hline
\endhead

\hline
\endfoot

\hline
\endlastfoot
AK Sco & 12700160 &  & 253.686667 & -36.888611 & PS & 12700160\_0 \\
AS 310 &  & 25733376 & 278.338333 & -4.968333 & P & 25733376\_0 \\
AS 470 & 12683008 &  & 324.059167 & 57.358611 & NF & 12683008\_2 \\
BD+30 549 & 14121472 & 14121472 & 52.3325 & 31.415833 & P & 14121472\_1 & \ding{51} \\
BF Ori & 18835968 & 5638144 & 84.305417 & -6.583611 & PS & 5638144\_0 \\
BH Cep & 21886720 &  & 330.42875 & 69.743333 & PS & 21886720\_0 \\
BO Cep & 21886976 &  & 334.225417 & 70.0625 & P & 21886976\_0 \\
BP Psc & 21814016 &  & 350.602917 & -2.228333 & NF & 21814016\_0 \\
CO Ori & 21870336 &  & 81.909583 & 11.4275 & S & 21870336\_0 \\
CPM 25 & 25736192 &  & 95.984583 & 14.507778 & PS & 25736192\_0 & \ding{51} \\
CQ Tau & 21875712 &  & 83.99375 & 24.748333 & PS & 21875712\_0 \\
DG Cir & 16828160 & 16828160 & 225.849167 & -63.383056 & S & 16828160\_0 \\
DK Cha & 12679168 & 22349312 & 193.32125 & -77.119722 & NF & 22349312\_5 \\
GSC 3975-0579 & 12683008 &  & 324.535417 & 57.446667 & P & 12683008\_6 \\
GSC 5360-1033 & 25735424 &  & 89.45625 & -14.092778 & P & 25735424\_0 \\
HBC 217 & 12675328 &  & 100.175833 & 9.560278 & P & 12675328\_0 \\
HBC 222 & 21880832 &  & 100.213333 & 9.746111 & P & 21880832\_0 \\
HBC 334 & 25731328 &  & 34.127917 & 55.383333 & P & 25731328\_0 & \ding{51} \\
HBC 442 & 18832640 &  & 83.559167 & -5.615 & PS & 18832640\_0 \\
HBC 717 & 21885696 &  & 313.025 & 44.287778 & S & 21885696\_0 \\
HD 101412 & 5640960 & 5640960 & 174.935 & -60.174444 & P & 5640960\_0 \\
HD 104237 & 12677632 & 12677632 & 180.020417 & -78.193056 & PS & 12677632\_0 \\
HD 130437 &  & 16826368 & 222.709167 & -60.286111 & P & 16826368\_0 \\
HD 132947 & 5643008 & 5643008 & 226.233333 & -63.131389 & NF & 5643008\_0 \\
HD 135344B &  & 5657088 & 228.951667 & -37.154444 & P & 5657088\_0 \\
HD 141926 & 25739264 &  & 238.590833 & -55.328889 & NF & 25739264\_0 \\
HD 142527 & 11005696 & 11005696 & 239.174583 & -42.323333 & PS & 11005696\_0 \\
HD 143006 & 5197568 & 9777152 & 239.65375 & -22.954444 & PS & 5197568\_0 \\
HD 149914 & 11000832 & 11000832 & 249.619167 & -18.220556 & NF & 11000832\_2 \\
HD 150193 &  & 11006208 & 250.074583 & -23.895833 & PS & 11006208\_0 \\
HD 155448 & 11006464 & 11006464 & 258.245 & -32.242778 & P & 11006464\_0 \\
HD 163296 &  & 5650944 & 269.08875 & -21.956111 & PS & 5650944\_0 \\
HD 17081 &  & 10998272 & 41.030417 & -13.858889 & NF & 10998272\_1 \\
HD 174571 & 25740544 &  & 282.696667 & 8.702778 & NF & 25740544\_0 \\
HD 179218 &  & 11006976 & 287.797083 & 15.7875 & P & 11006976\_0 \\
HD 200775 & 16207104 & 16207104 & 315.40375 & 68.163333 & S & 16207104\_0 \\
HD 244604 & 11001344 & 11001344 & 82.98875 & 11.294722 & PS & 11001344\_0 \\
HD 249879 & 25735680 &  & 89.7325 & 16.665833 & PS & 25735680\_0 \\
HD 250550 & 16826624 & 16826624 & 90.5 & 16.515833 & PS & 16826624\_0 \\
HD 259431 &  & 11003392 & 98.271667 & 10.322222 & P & 11003392\_0 \\
HD 288012 & 21889280 &  & 83.27 & 2.469444 & PS & 21889280\_0 \\
HD 290380 & 21889024 &  & 80.879167 & -1.073333 & PS & 21889024\_0 \\
HD 290764 & 21890304 &  & 84.522083 & -1.256111 & P & 21890304\_0 \\
HD 290770 & 25734656 &  & 84.26 & -1.6225 & PS & 25734656\_0 \\
HD 319896 & 25739776 &  & 262.774583 & -35.141389 & HAC & 25739776\_0 & \ding{51} \\
HD 35929 & 10998528 & 10998528 & 81.928333 & -8.3275 & NF & 10998528\_2 \\
HD 36112 & 11001088 & 11001088 & 82.614583 & 25.3325 & PS & 11001088\_0 \\
HD 36408 & 25734400 &  & 83.05875 & 17.058056 & NF & 25734400\_0 \\
HD 36917 & 11001600 & 11001600 & 83.695833 & -5.570833 & P & 11001600\_0 \\
HD 37258 & 18814720 & 10998784 & 84.247083 & -6.154444 & S & 18814720\_0 \\
HD 37357 & 11001856 & 11001856 & 84.44625 & -6.708333 & PS & 11001856\_0 \\
HD 37806 & 11002368 & 11002368 & 85.259583 & -2.716944 & PS & 11002368\_0 \\
HD 38087 & 11002624 & 11002624 & 85.7525 & -2.3125 & P & 11002624\_0 \\
HD 38120 & 11002880 & 11002880 & 85.799583 & -4.997222 & PS & 11002880\_0 \\
HD 39014 &  & 10999296 & 86.192917 & -65.735556 & NF & 10999296\_1 \\
HD 46060 & 25732864 &  & 97.7075 & -9.654167 & NF & 25732864\_0 & \ding{51} \\
HD 50083 & 11000064 & 11000064 & 102.940833 & 5.084444 & NF & 11000064\_1 \\
HD 50138 &  & 11003648 & 102.889167 & -6.966389 & PS & 11003648\_0 \\
HD 53367 & 16826880 & 16826880 & 106.10625 & -10.454444 & NF & 16826880\_0 \\
HD 56895B & 11003904 & 11003904 & 109.6325 & -11.192778 & NF & 11003904\_0 \\
HD 58647 & 11004160 & 11004160 & 111.48375 & -14.178889 & NF & 11004160\_0 \\
HD 59319 & 25736704 &  & 112.153333 & -21.963611 & NF & 25736704\_0 \\
HD 72106B & 11004416 & 11004416 & 127.395417 & -38.605833 & PS & 11004416\_0 \\
HD 76534 & 16827136 & 16827136 & 133.78625 & -43.466667 & NF & 16827136\_0 \\
HD 85567 & 11004672 & 11004672 & 147.61875 & -60.9675 & PS & 11004672\_0 \\
HD 95881 & 11004928 & 11004928 & 165.49 & -71.513333 & PS & 11004928\_0 \\
HD 96042 & 14206464 &  & 165.91875 & -59.433056 & NF & 14206464\_0 \\
HD 9672 & 4928768 & 4928768 & 23.657917 & -15.676389 & NF & 4928768\_0 \\
HD 97048 & 12697088 & 12697088 & 167.013333 & -77.654722 & P & 12697088\_1 \\
HD 98922 &  & 5640704 & 170.632083 & -53.369722 & PS & 5640704\_0 \\
HR 5999 &  & 11005952 & 242.142917 & -39.105278 & PS & 11005952\_0 \\
Hen 2-80 &  & 25738496 & 185.596667 & -63.288056 & P & 25738496\_0 \\
Hen 3-1191 &  & 16828928 & 246.812917 & -48.6575 & P & 16828928\_0 \\
Hen 3-847 &  & 25733120 & 195.324167 & -48.888611 & PS & 25733120\_0 \\
IL Cep & 25734144 &  & 343.315 & 62.145833 & PS & 25734144\_0 \\
KK Oph &  & 16827392 & 257.53375 & -27.255278 & PS & 16827392\_0 \\
LKHa 260 & 25747456 &  & 274.789167 & -13.844722 & S & 25747456\_0 \\
LKHa 338 & 21880064 &  & 92.69625 & -6.214167 & PS & 21880064\_0 \\
LkHa 215 & 14124032 & 14124032 & 98.174167 & 10.159444 & P & 14124032\_0 & \ding{51} \\
LkHa 257 & 25733888 &  & 328.578333 & 47.202778 & P & 25733888\_0 \\
LkHa 339 & 25732352 &  & 92.740833 & -6.244444 & PS & 25732352\_0 \\
MWC 1080 & 21887488 &  & 349.356667 & 60.845278 & P & 21887488\_0 \\
MWC 137 &  & 26897920 & 94.689583 & 15.281111 & P & 25732608\_0 \\
MWC 314 & 27569664 &  & 290.391667 & 14.8825 & NF & 27569664\_0 \\
MWC 593 & 25746432 &  & 267.2925 & -24.239167 & NF & 25746432\_0 & \ding{51} \\
MWC 623 &  & 22902016 & 299.13125 & 31.105556 & B & 22902016\_0 \\
MWC 657 & 22903552 & 22903552 & 340.674167 & 60.400278 & S & 22903552\_0 \\
MWC 878 &  & 25739520 & 261.18625 & -38.730833 & S & 25739520\_0 \\
MWC 930 & 25444352 &  & 276.605 & -7.221667 & NF & 25444352\_0 \\
NV Ori & 21876480 &  & 83.880833 & -5.5525 & PS & 21876480\_0 \\
NX Pup & 21882112 & 16827904 & 109.867917 & -44.586389 & S & 16827904\_0 \\
PDS 022 & 25735936 &  & 90.904583 & -14.884167 & PS & 25735936\_0 \\
PDS 211 & 22894848 & 22894848 & 92.572083 & 29.421389 & PS & 22894848\_0 \\
PDS 241 &  & 25745408 & 107.161667 & -4.318056 & P & 25745408\_0 \\
PDS 27 &  & 25736448 & 109.899583 & -17.655 & NF & 25736448\_0 \\
PDS 344 & 25738241 &  & 175.136667 & -64.535 & P & 25738241\_0 & \ding{51} \\
PDS 37 & 25447169 & 25737473 & 152.50125 & -57.035278 & PS & 25737473\_0 \\
PDS 415N & 12674304 & 12674304 & 244.655 & -24.088333 & P & 12674304\_1 \\
PDS 477 & 25740288 &  & 270.12625 & -16.790556 & PS & 25740288\_0 \\
PDS 543 & 25746688 & 25746688 & 282.002917 & 2.904722 & S & 25746688\_0 \\
PDS 581 &  & 25740800 & 294.07875 & 29.547222 & NF & 25740800\_0 \\
PDS 69 &  & 25739008 & 209.432917 & -39.979722 & PS & 25739008\_0 \\
PX Vul & 21884928 &  & 291.667917 & 23.8975 & PS & 21884928\_0 \\
RR Tau & 5638400 & 5638400 & 84.877083 & 26.374167 & P & 5638400\_0 \\
RY Ori & 21871616 &  & 83.04125 & -2.829722 & S & 21871616\_0 \\
SAO 185668 & 11006720 & 11006720 & 265.981667 & -22.095833 & NF & 11006720\_0 \\
SAO 220669 & 25737216 &  & 133.94125 & -44.420556 & HAC & 25737216\_0 & \ding{51} \\
V1012 Ori & 25731584 &  & 77.902083 & -2.38 & P & 25731584\_0 \\
V1295 Aql & 11007232 & 11007232 & 300.760417 & 5.738056 & S & 11007232\_0 \\
V1686 Cyg &  & 16827648 & 305.122083 & 41.357778 & P & 16827648\_0 \\
V1787 Ori & 18834176 &  & 84.53875 & -6.821389 & PS & 18834176\_0 \\
V1818 Ori &  & 25735168 & 88.4275 & -10.400278 & S & 25735168\_0 \\
V1977 Cyg & 21885184 &  & 311.90625 & 43.790278 & NF & 21885184\_0 \\
V346 Ori & 16265216 &  & 81.178333 & 1.73 & P & 16265216\_0 \\
V373 Cep &  & 25733632 & 325.778333 & 66.115 & P & 25733632\_0 \\
V380 Ori &  & 25731840 & 84.105833 & -6.716111 & PS & 25731840\_0 \\
V388 Vel & 18596864 & 18596864 & 130.572083 & -40.736111 & PS & 18596864\_0 \\
V590 Mon & 12674816 & 12674816 & 100.185833 & 9.800556 & PS & 12674816\_1 \\
V594 Cas &  & 25731072 & 10.82625 & 61.911111 & PS & 25731072\_0 \\
V669 Cep &  & 22903040 & 336.66125 & 61.225556 & S & 22903040\_0 \\
V892 Tau &  & 3869696 & 64.669167 & 28.320833 & B & 3869696\_0 \\
V921 Sco &  & 4898048 & 254.778333 & -42.702222 & P & 4898048\_0 \\
VV Ser & 5651200 & 5651200 & 277.199583 & 0.144444 & PS & 5651200\_0 \\
WW Vul & 16828672 & 16828672 & 291.495 & 21.208611 & PS & 16828672\_0 \\ \hline
\end{longtable}

\newpage
\bibliography{ms2025-0088.bib}
\bibliographystyle{mnras}
\label{lastpage}

\end{document}